\documentclass[aps,reprint,amsmath,amssymb,nofootinbib]{revtex4-1}
\usepackage{xcolor}
\usepackage{graphicx} 
\usepackage[caption=false]{subfig}

\usepackage{gensymb} 
\usepackage{amsmath} 
\usepackage{braket} 
\usepackage{booktabs} 
\usepackage{ulem}
\usepackage[colorlinks=true]{hyperref} 
\hypersetup{colorlinks,
	citecolor=blue
}
\usepackage[capitalize]{cleveref}

\newcommand{\ba}{\begin{eqnarray}}
\newcommand{\ea}{\end{eqnarray}}

\begin{document}


\title{The SuSAv2 model for inelastic neutrino-nucleus scattering}


\author{J. Gonzalez-Rosa$^a$, G. D. Megias$^{a,b}$, J. A. Caballero$^{a,c}$, M. B. Barbaro$^{d,e,f}$}

\affiliation{$^a$Departamento de F\'isica at\'omica, molecular y nuclear, Universidad de Sevilla, 41080 Sevilla, Spain}

\affiliation{$^b$ Research Center for Cosmic Neutrinos, Institute for Cosmic Ray Research, University of Tokyo, Kashiwa, Chiba 277-8582, Japan}

\affiliation{$^c$ Instituto de F\'{\i}sica Te\'orica y Computacional Carlos I, Granada 18071, Spain}

\affiliation{$^d$Dipartimento di Fisica, Universit\'a di Torino, 
10125 Torino, Italy}

\affiliation{$^e$ INFN, Sezione di Torino, 
10125 Torino, Italy}

\affiliation{$^f$ 
IPSA Paris, 
94200 Ivry-sur-Seine, France}




\date{\today}

\begin{abstract}

	
The susperscaling model SuSAv2, already available for charged-current neutrino-nucleus cross sections in the quasielastic region, is  extended to the full inelastic regime. In the model  the resonance production and deep inelastic reactions are described through the extension to the neutrino sector of the  SuSAv2 inelastic model developed for ($e,e'$) reactions, which combines phenomenological structure functions with a nuclear scaling function. 
This work also compares  two different descriptions  of the $\Delta$ resonance region, one based on a global scaling function for the full inelastic spectrum and the other on a semi-phenomenological $\Delta$ scaling function extracted from  ($e,e'$) data for this specific region and updated with respect to previous work.
The results  of the model are tested against ($e,e'$) data on $^{12}$C, $^{16}$O, $^{40}$Ca and $^{40}$Ar and applied to the study of the charged current inclusive neutrino cross-section on $^{12}$C and $^{40}$Ar measured by the T2K, MicroBooNE, ArgoNEUT and MINERvA experiments, thus covering several kinematical regions. 
		
\end{abstract}


\maketitle

\section{Introduction \label{Introduction}}

The enormous effort made in recent years in the development of neutrino oscillation experiments has motivated many theoretical analyses devoted to obtain accurate descriptions of neutrino-nucleus reactions. These are needed in order to reduce one of the leading experimental uncertainties, that associated to nuclear effects, for the determination of oscillation parameters and the violation of the charge-parity symmetry in the neutrino sector~\cite{NuSTEC:2017hzk}.
Many neutrino experiments~\cite{Abratenko_2019,filkins_double-differential_2020,Abe_2018,Acciarri_2014,AguilarArevalo:2010zc,NOvA:2019cyt,dunecollaboration2016longbaseline,protocollaboration2018hyperkamiokande} (MiniBooNE, MicroBooNE, T2K, NOvA, MINERvA, ArgoNEUT and, in future,  DUNE and HyperK) operate in the 0.5-10 GeV region, where several mechanisms contribute to the nuclear response: from the excitation of collective states at the lowest  transferred energies up to the deep inelastic scattering (DIS) process at the highest kinematics, embracing also the quasi-elastic (QE) regime, associated to one-nucleon knockout, the emission of two nucleons, commonly denoted as two-particle-two-hole (2p2h) channel, and the resonance region, corresponding to the excitation of nucleon resonances followed by their decay and associated production of pions and other mesons.

 Most of these experiments are focused on the measurements of CC0$\pi$ (or ``quasielastic-like") events, which are defined as charged-current (CC)
 interactions,  characterized by having no pions ($0\pi$) detected in the final state. These 
 events are dominated by the QE and 2p2h channels and, in accordance, many theoretical studies of these contributions have been carried out in recent years~\cite{Nieves12b,Nieves11,Martini10,Martini12,Gallmeister16,Meucci11a,Meucci14,Rocco16,Lovato16,Pandey16,Gonzalez-Jimenez19,Gonzalez-Jimenez20}. 
 The inelastic region, corresponding to resonant and non-resonant meson production and DIS,  is of relevance in the  CC-inclusive process, where all reaction channels, including inelasticities, are considered. It also plays a role in the  experimental analysis of CC0$\pi$ measurements, where it can represent an important background \cite{athar_neutrino_2020,minerva2019}. 
 However, the nuclear models are not yet as well developed for the inelastic region as for the quasi-elastic one and current theoretical efforts on this way are being carried out~\cite{PhysRevD.97.013004,PhysRevC.100.045501,PhysRevC.80.065501,Vagnoni:2017hll,ivanov_superscaling_2012,Singh:1998ha,PhysRevC.79.044603,PhysRevD.92.074024,Barbaro:2021psv}. 

Within the inelastic regime, the excitation of nucleon resonances is one of the most relevant channels. It is related to larger energy transfers than the one corresponding to the QE process, and involves higher hadronic invariant masses in the final state. The most important contribution comes from the $\Delta(1232)$ resonance.
Preliminary studies carried out by some of the authors of this paper have been presented in previous works \cite{PhysRevC.71.015501,2004_inelastic,ivanov_charged-current_2016}, but contributions associated to nuclear effects still need further investigation.
Also, there is a lack of knowledge about heavier resonances which belong to the so-called Shallow Inelastic Scattering (SIS) region, i.e. the transition region between resonance excitations and DIS. 
The SIS region can also contribute significantly to the determination of neutrino oscillation parameters as it can be relevant for both signal and background estimates. 
The analysis of the resonance form factors and inelastic structure functions of the nucleons mainly comes from the study of electron scattering data, which introduces some limitations in the neutrino sector due to the missing axial response in electron-nucleus reactions. This requires relying on different approaches, such as QCD calculations, 
quark models and Parton Distribution Functions (PDFs) or phenomenological models~\cite{athar_neutrino_2020,gluck_dynamical_1998,bodek_axial_2013,PhysRevD.24.1400,PhysRevC.81.055213,PhysRevC.77.065206}. However, most of these approaches are affected by large uncertainties and some kinematical limitations, which make difficult to obtain a consistent description of the inelastic regime. The SIS and DIS regions are thus a subject of continuing study where more experimental and theoretical efforts are required.

Several measurements of CC inclusive  neutrino-nucleus scattering cross sections have been performed recently by various experiments. In the case of T2K \cite{t2k_collaboration_measurement_2014,Abe_2018}, MicroBooNE~\cite{Abratenko_2019} and SciBooNE \cite{sciboone_collaboration_measurement_2011} the neutrino energy is peaked around 0.6 GeV, which makes 
quasielastic scattering, one pion production and 2p2h excitations  the main contributions to the cross section, being the QE regime the dominant one.
 On the other hand, other recent experiments, such as MINERvA, NOvA or ArgoNEUT and the future DUNE~\cite{Filkins_2020, NOvA:2019cyt, argoneut_collaboration_first_2018, dunecollaboration2016longbaseline}, present a significant number of events at energies higher than 3 GeV where inelasticities play an important role. With increasing neutrino energies, an accurate description of the inelastic spectrum is becoming more and more important, thus motivating the development of new analyses of the nucleon structure functions and sophisticated nuclear models to address these contributions and reduce the large uncertainties in current studies.  

Several models have been developed to describe the inelastic region, mainly pion production in nuclei, which provide different treatments of the initial nuclear state, the production of pions on a bound nucleon, and the interaction of the pions and nucleons in the residual nucleus. Although some older studies are  based on the Fermi gas  of non-interacting nucleons~\cite{Kim:1996bt,Singh:1998ha}, recently different groups have developed more sophisticated descriptions that incorporate the relativistic mean field theory~\cite{PhysRevD.97.013004,PhysRevC.100.045501,PhysRevC.79.044603}, Random Phase Approximation calculations~\cite{PhysRevC.80.065501} or spectral functions~\cite{Benhar:2005dj,Vagnoni:2017hll}.

In this work we extend the SuperScaling model SuSAv2,   previously applied to the study of  quasielastic neutrino scattering and briefly summarized in Sect.II,  to the inelastic regime, following what has been done in  Refs.~\cite{barbaro_inelastic_2004,maieron_superscaling_2009,Megias_2016} for electron scattering.  In the case of neutrino reactions the main difficulty  arises from the poor knowledge of the weak inelastic structure functions, in particular the axial one $W_3$, across the full inelastic spectrum. In the present study we explore two different options. The first one focusses on the $\Delta$ resonance, for which the  experimental information on the weak  form factors is better established, and combines the elementary cross section $\nu+N\to l +\Delta$ ($N$ being the hit nucleon and $l$ the outgoing lepton) with a semi-phenomenological scaling function $f^\Delta$ to be used only  in the $\Delta$ region. The function $f^\Delta$ is extracted from the analysis of inclusive  electron scattering on $^{12}$C by subtracting from the data the QE and 2p2h contributions evaluated in the SuSAv2 model; it carries information on the nuclear dynamics in this region and on the propagation of the resonance in the medium. This approach was taken in Refs.~\cite{PhysRevC.71.015501,ivanov_charged-current_2016} and is now revisited using an updated version of the model for the QE and 2p2h regions. The resulting model, labeled as "SuSAv2-$\Delta$",  is presented in Sect.~III.
The second option is the extension of the SuSAv2 model to the complete inelastic spectrum - resonant, non-resonant and DIS - and represents a generalization of what has been done in  the case of electrons in Ref.~\cite{barbaro_inelastic_2004}. This model 
will be referred to as "SuSAv2-inelastic" and is described in detail in Sect.~IV together with an analysis of different inelastic weak structure functions.
In Sect.~V we compare both the SuSAv2-$\Delta$ and SuSAv2-inelastic models with electron scattering data as a solid benchmark to test their validity before their application to the neutrino sector. In Sect.~VI we show a comparison of the previous models with CC neutrino-nucleus scattering data from several experiments, different nuclei, and at different kinematics. In Sect.~VII we draw our conclusions.





\section{The SuperScaling approach}



The SuperScaling Approach (SuSA), based on the superscaling properties exhibited by inclusive electron scattering~\cite{donnelly_superscaling_1999}, has been successfully applied to the analysis of both electron and neutrino cross sections~\cite{Megias_2016,Megias16b,Megias18,Megias_2019} for several nuclei. This model, which was originally developed as a semi-phenomenological approach for the QE and $\Delta$-resonance regions~\cite{PhysRevC.71.015501}, was subsequently extended to the full inelastic regime~\cite{maieron_superscaling_2009}
 for electron-nucleus reactions. The model was later updated (SuSAv2) using Relativistic Mean Field (RMF) ingredients to develop a theory-based approach  to the QE  regime~\cite{Gonzalez-Jimenez:2014eqa} and including a fully relativistic calculation of 2p2h contributions~\cite{Amaro:2010sd,RuizSimo:2016rtu},   improving the description of nuclear effects and the agreement with data~\cite{Megias_2016}.
  
Although the detailed description of the SuperScaling model and its formalism can be found in several references -- see \cite{Amaro_2020,Amaro:2021sec} for recent reviews -- here we recall its main features before introducing the new ingredients of the model.

The SuSA model has its foundations in the analysis of the inclusive electron-nucleus cross sections. In the QE region, it has been observed that the global ($e,e'$) inclusive cross section data exhibit a general independence of the transferred momentum (scaling of 1st kind) and of the nuclear species (scaling of 2nd kind) when divided by the single-nucleon cross section and multiplied by the corresponding Fermi momentum $k_F$. The simultaneous occurrence of both kinds of scaling is called superscaling. The above-mentioned ratio defines a scaling function $f^{QE}(\psi)$ of the scaling variable $\psi=\psi(\omega,q)$, given by the following combination of the energy $(\omega)$ and momentum $(q)$ transferred to the nucleus~\cite{Donnelly:1998xg}:
\begin{equation}
    \psi=\frac{1}{\sqrt{\xi_{F}}}\frac{\lambda - \tau}{\sqrt{(1 + \lambda)\tau + \kappa\sqrt{(1 + \tau)\tau}}} \ ,
    \label{eq:psi}
\end{equation}
where the dimensionless transferred momentum  ($\kappa \equiv q/2m_{N}$), energy ($\lambda \equiv \omega/2m_{N}$) and four-momentum $\tau \equiv \kappa^{2}-\lambda^{2}$ ($\tau=Q^{2}/4m^{2}_{N}$) and the dimensionless Fermi kinetic energy ($\xi_{F} \equiv \sqrt{1 + (k_{F}/m_{N})^{2}} -1$) have been introduced in terms of the nucleon mass $m_N$.
 By definition, this scaling function embeds most of the nuclear dynamics and thus it can be extended from electron to neutrino reactions. 

Since some longitudinal  and transverse (with respect to the momentum transfer $\bf q$) separated $(e,e')$ data exist in the QE regime~\cite{Jourdan:1996np}, based on the Rosenbluth separation, one can also define and  analyze the longitudinal and transverse  scaling functions, defined as
\begin{equation}
    f^{QE}_{L,T}=k_{F}\frac{R^{QE}_{L,T}}{G^{QE}_{L,T}}\ ,
    \label{eq:fl}
\end{equation}
where  $R^{QE}_{L,T}$ are the longitudinal and transverse nuclear responses and $G^{QE}_{L,T}$  the corresponding single-nucleon responses (see \cite{Barbaro:1998gu}  for their explicit expressions). The data analysis performed in \cite{donnelly_superscaling_1999} shows that in the QE peak the longitudinal scaling function \eqref{eq:fl}  superscales, i.e. it only depends on a single variable $\psi$, for all nuclei and most kinematics except for very low densities and momenta (roughly $q\le$ 300 MeV/c). On the contrary, the transverse scaling function 
 exhibits some scaling violations due to other contributions that can play a significant role in the QE region and that are mainly transverse, such as 2p2h and $\Delta$ resonance. In its first version~\cite{PhysRevC.71.015501} the SuSA approach assumed $f^{QE}_L=f^{QE}_T=f^{QE}$ (scaling of 0-th kind), namely that the superscaling  function was the same in the longitudinal and transverse channels, and applied a phenomenological fit of the
longitudinal  ($e,e'$) data to construct the model. 
 Later, this phenomenological description was improved using the microscopic RMF model, as it was observed that the theoretical scaling functions derived from this theory matched with the ($e,e'$) scaling data. 
 In this microscopic approach (``SuSAv2") the RMF scaling functions are employed to describe both electron and neutrino reactions~\cite{Gonzalez-Jimenez:2014eqa}. The SuSAv2 model has the merit of reproducing both the height and the 
shape of the 
longitudinal scaling function while predicting a slight enhancement of the transverse scaling function, which is supported by the separate L/T data analysis~\cite{Jourdan:1996np} and related to the relativistic nature of the RMF model. In the RMF theory the initial nucleon's wave function is a bound solution of the Dirac equation in presence of two strong scalar and vector relativistic potentials, while the final nucleon is a scattering eigenstate of the same Hamiltonian: as a consequence Pauli blocking and binding energy effects are intrinsically taken into account in this model~\cite{Gonzalez-Jimenez19}. The unrealistically strong effect of the RMF potentials at high kinematics, where the 
distorsion of the nucleon wave function due to final state interactions (FSI) should instead disappear and the Relativistic Plane Wave Impulse Approximation (RPWIA) limit be recovered, is corrected in the SuSAv2 model by implementing a transition function between the RMF and RPWIA quasielastic scaling functions \cite{megias_charged-current_2017,Megias_2016}.

The extension of the SuperScaling approach to the inelastic regime is based on the assumption that a scaling function $f=f^{QE}$ can be used also at higher energy transfers to describe the nuclear dynamics: the corresponding nuclear responses are obtained by folding $f$ with the appropriate elementary structure functions. 
%
%
In particular, the SuperScaling Approach has been applied to the analysis of the full inelastic regime in the case of electron scattering in \cite{2004_inelastic,maieron_superscaling_2009}, where phenomenological electromagnetic structure functions, $W_1$ and $W_2$, have been used to describe the elementary inelastic processes. This approach has been recently improved using the RMF theory (SuSAv2-inelastic model~\cite{megias_charged-current_2017})  in the QE region. Moreover,  a semi-phenomenological treatment of the $\Delta$-resonance region has been developed within the SuSA model for both electron and neutrino reactions (SuSA-$\Delta$ model~\cite{PhysRevC.71.015501,ivanov_charged-current_2016}). 
In the next sections we present an improved, more accurate version of the semi-phenomenological SuSA-$\Delta$ model and an extension of the SuSAv2-inelastic model to the neutrino sector.



Finally, in order to describe the full spectrum, 2p2h excitations induced by Meson Exchange Currents (MEC) have also been implemented in the SuSAv2 model for both electron and neutrino scattering using the results of the microscopic calculations \cite{pace_2p2h_2003,RuizSimo:2016rtu,PhysRevD.91.073004,PhysRevD.90.033012,PhysRevD.91.073004}. 


\section{The SuSAv2-$\Delta$ model}\label{susav2delta}

In this section we introduce a model to describe the electron- and CC neutrino-nucleus reactions associated to the $\Delta$ excitation based on the SuperScaling approach described in the previous section. This approach allows one to handle the QE and $\Delta$ regions in a unified framework and can be applied to high energies due to its relativistic nature. 

The idea of using SuperScaling to model the $\Delta$-resonance region in neutrino-nucleus scattering was proposed in Ref.~\cite{PhysRevC.71.015501} and further developed and compared with data in Ref.~\cite{ivanov_charged-current_2016}. In particular it was shown that the residual strength  obtained after subtracting the QE and 2p2h contributions from the experimental $(e,e')$ cross section measured at different kinematics displays a scaling behaviour similar to the one observed in the quasielastic channel, provided a new scaling variable associated to the $\Delta$ production is introduced:
\begin{equation}
    \psi_{\Delta}=\frac{1}{\sqrt{\xi_{F}}}\frac{\lambda - \tau \rho_{\Delta}}{\sqrt{(1 + \lambda\rho_{\Delta})\tau + \kappa\sqrt{\tau(1 + \tau\rho^{2}_{\Delta})}}}\ ,
\label{eq:psid}
\end{equation}
where
\begin{equation}\label{rho_mu}
    \rho_{\Delta}=1 + \frac{1}{4\tau}(\mu^{2}_{\Delta}  - 1)
\end{equation}
is the inelasticity parameter and 
\begin{equation}
    \mu_{\Delta}=\frac{m_{\Delta}}{m_{N}}
\end{equation}
the dimensionless $\Delta$ mass. The $\Delta$ scaling variable \eqref{eq:psid} vanishes when the energy transfer corresponds to the excitation of a $\Delta$ resonance on a free nucleon at rest $(\omega=\sqrt{q^2+m_\Delta^2} -m_N$), which coincides with  the center of the $\Delta$-resonance peak, and it  reduces to the quasielastic scaling variable \eqref{eq:psi} for $m_\Delta\to m_N$. More specifically, by dividing the cross section
\begin{widetext}
\begin{equation}
\label{cs-only-delta0}
    \biggl(\frac{d^{2}\sigma}{d\Omega_{e}d\omega}\biggr)^{\Delta} \equiv \biggl(\frac{d^{2}\sigma}{d\Omega_{e}d\omega}\biggr)^{exp.} 
     -\biggl(\frac{d^{2}\sigma}{d\Omega_{e}d\omega}\biggr)^{SuSA-QE}   - \biggl(\frac{d^{2}\sigma}{d\Omega_{e}d\omega}\biggr)^{2p2h}\,,
          \end{equation}
\end{widetext}
where the QE and 2p2h contributions are evaluated using the SuSA model, 
by the elementary $N\to\Delta$ cross section~\cite{PhysRevC.71.015501},  one obtains a function
 \begin{equation}
\label{f-delta1}
f^{\Delta}(\psi_{\Delta})=k_{F}\frac{\biggl(\frac{d^{2}\sigma}{d\Omega_{e}d\omega}\biggr)^{\Delta}}{\sigma_{Mott}(\upsilon_{L}G^{\Delta}_{L} + \upsilon_{T}G^{\Delta}_{T})}\ ,
\end{equation}
which approximately depends only on the variable $\psi_\Delta$. This result indicates that this region is dominated by the $\Delta$ resonance excitation and that the nuclear effects acting in this regime can be effectively embodied in a scaling function. However, this $\Delta$ scaling is valid only for $\psi_\Delta \lesssim 0$, while in the right part of the peak it is broken due to the opening of higher inelastic channels, namely the higher resonances (HR) and DIS contributions.

In this work we revisit this procedure by using the last version of the SuSAv2-QE and 2p2h models. A further substantial improvement of the model consists in employing the SuSAv2-inelastic model (see next Section)  to remove from the inclusive electron scattering data contributions beyond the $\Delta$ resonance, namely the HR and  DIS channels, in such a way that the $\Delta$-resonance peak is better isolated. 
 Accordingly, we modify Eq.~\eqref{cs-only-delta0} as follows
%
\begin{widetext}
\begin{equation}
\label{cs-only-delta1}
    \biggl(\frac{d^{2}\sigma}{d\Omega_{e}d\omega}\biggr)^{\Delta} \equiv \biggl(\frac{d^{2}\sigma}{d\Omega_{e}d\omega}\biggr)^{exp.} 
     -\biggl(\frac{d^{2}\sigma}{d\Omega_{e}d\omega}\biggr)^{SuSAv2-QE}   - \biggl(\frac{d^{2}\sigma}{d\Omega_{e}d\omega}\biggr)^{2p2h}- \biggl(\frac{d^{2}\sigma}{d\Omega_{e}d\omega}\biggr)^{HR + DIS}
     \end{equation}
\end{widetext}
%
and, using Eq.~\eqref{f-delta1}, we obtain an improved $\Delta$ scaling function. 

 
  \begin{figure*}[!htbp]
    \centering
    \includegraphics[width=1.0\textwidth]{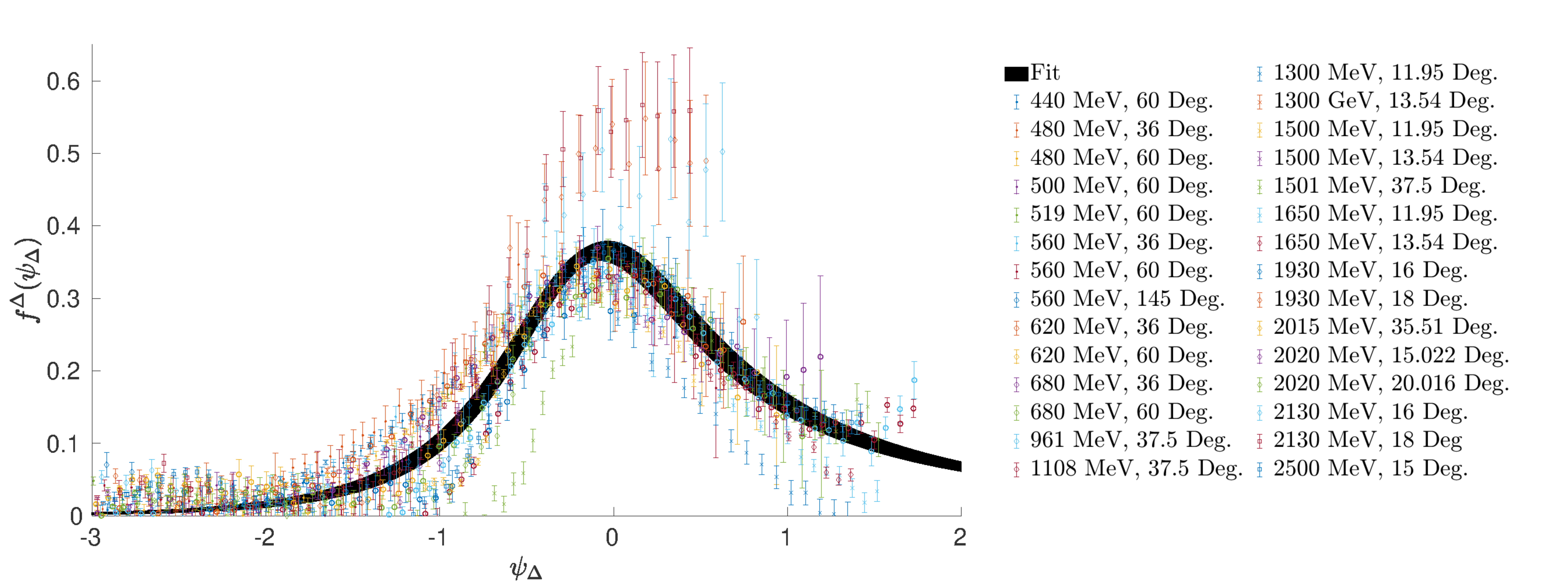}
    \caption{\label{scaling1}  Experimental values of $f^{\Delta} (\psi_{\Delta})$ 
 together with a phenomenological fit of the $\Delta$ scaling function. Data are labeled by the corresponding electron beam energy and scattering angle.}
\end{figure*} 
 
In fig.~\ref{scaling1} we show the semi-phenomenological scaling function obtained by using Eqs. \eqref{f-delta1} and  \eqref{cs-only-delta1} applied to $^{12}$C data from \cite{benhar_archive_2006}. 
When we  compare with the previous $\Delta$ scaling function obtained in \cite{ivanov_charged-current_2016}, as expected,  a similar behavior is shown at negative $\psi_\Delta$ values (below the $\Delta$ peak). However, at kinematics above the $\Delta$ peak ($\psi_\Delta>0$) there is a reduction of $f^\Delta$ in the new approach due to the subtraction of the HR+DIS contributions. 

\section{SuSAv2-inelastic model \label{SuSav2-inel}}

While the above described model is restrained to two particular reaction mechanisms, i.e., $\Delta$ production and QE scattering, in this section we extend the 
formalism based on the SuperScaling Approach and the RMF theory to the full inelastic regime. 

Following previous studies on the inelastic RFG and scaling modeling~\cite{2004_inelastic}, an extension of the SuSAv2-QE formalism to  the complete inelastic spectrum - resonant, non-resonant and deep inelastic scattering -  has  been proposed in~\cite{Megias_2016} for the analysis of electron reactions. This was carried out by employing phenomenological fits to the single-nucleon inelastic structure functions together with an extension of the SuSAv2-QE scaling functions to the inelastic regime, yielding a very good agreement with inclusive electron scattering data at high energies. Here we extend this description to the neutrino sector.

Although the general formalism describing inclusive inelastic lepton-nucleus reactions within the SuSAv2 approach has been presented in \cite{Megias_2016}, here we summarize its main features, emphasizing its extension to the neutrino case.

The expression for the CC double differential neutrino cross section is given by
\begin{equation}
    \frac{d^{2}\sigma}{d\Omega_{l} d\omega_{l}}=\frac{G^{2}_{F}p_{l}}{4\pi^{2}E_{\nu}}L_{\mu\nu}W^{\mu\nu},
\end{equation}
being $L_{\mu\nu}$ and $W^{\mu\nu}$ the leptonic and the inelastic hadronic tensor, respectively~\cite{Alberico97}. The term $p_l$ ($E_\nu$) refers to the outgoing lepton momentum (neutrino beam energy) and $G_F$ is the Fermi constant.


The inelastic nuclear responses are obtained by integrating the nuclear responses depending on the final-state invariant mass $W_X$ over all possible final hadronic states \cite{2004_inelastic,Megias_2016}
\begin{align}\label{inelastic_response}
    R^{inel}_{K}(\kappa, \tau)=\frac{N m^{3}_{N}}{k_{F}^{3}\kappa}\xi_{F}\int_{\mu^{min}_{X}}^{\mu^{max}_{X}} d\mu_{X}\mu_{X}f^{model}(\psi_{X})G^{inel}_{K},
\end{align}
being $N$ the number of nucleons, $\mu_X$ the dimensionless invariant mass
\begin{equation}
    \mu_{X}=\frac{W_{X}}{m_{N}} ,
\end{equation}
$G^{inel}_K$ the inelastic response of a single nucleon and $K$ an index related to the different longitudinal and transverse channels.

The inelastic nuclear responses are defined in terms of the different components of the hadron tensor $W^{\mu\nu}$, which is defined as \begin{align}\label{inelastic_response}
    W_{inel}^{\mu\nu}(\kappa, \tau)=\frac{N m^{3}_{N}}{k_{F}^{3}\kappa}\xi_{F}\int_{\mu^{min}_{X}}^{\mu^{max}_{X}} d\mu_{X}\mu_{X}f^{model}(\psi_{X})G_{inel}^{\mu\nu},
\end{align}
in the following way: 
\begin{eqnarray}\label{RCb}
R_{CC} =W^{00} \quad ; \quad  \\R_{CL} = -\frac12\left( W^{03}+W^{30} \right) \quad ; \quad\\ R_{LL}  = W^{33}  \quad ; \\
R_T =  W^{11}+W^{22} \quad ; \quad  \\ R_{T'} = -\frac{i}{2}\left( W^{12}-W^{21} \right)\,, \label{RTPb}
\end{eqnarray}
for the neutrino case, and
\begin{eqnarray}
R_{L} &=&W^{00} \label{RCb}\\
R_T &=&  W^{11}+W^{22} \,, \label{RTPb}
\end{eqnarray}
for electromagnetic interactions.
The expression for the inelastic nucleon tensor $G^{\mu\nu}$ in terms of the inelastic structure functions $W_i(\tau,\rho_X)$ is given by
\begin{widetext}
\begin{eqnarray} G^{\mu\nu}_{inel}(\kappa,\tau,\rho)&=&
-\left[W_1(\tau,\rho_X)+\frac{1}{2}W_2(\tau,\rho_X){\cal
    D}(\kappa,\tau,\rho_X)\right]
\left(g^{\mu\nu}+\frac{\kappa^\mu\kappa^\nu}{\tau}\right)
\nonumber\\
&+& W_2(\tau,\rho_X)\left[1+\tau\rho_X^2+\frac{3}{2}{\cal
    D}(\kappa,\tau,\rho_X)\right] \frac{a^\mu a^\nu}{\tau}\nonumber\\
&\mp& i W_3(\tau,\rho_X)\epsilon^{\mu\nu\alpha\beta}\left[\left(\frac{1}{2}(\epsilon_F+\epsilon_0)+\lambda\rho_X\right)\frac{a_\alpha\kappa_\beta}{\kappa}-\rho_X\kappa_\alpha\kappa_\beta\right]\ ,
\label{eq:Umunu}
\end{eqnarray} 
\end{widetext}
with $a^\mu=(\kappa,0,0,\lambda), \kappa^\mu=(\lambda,0,0,\kappa)$ and $\rho_X$ the inelasticity parameter defined as
\begin{equation}
 \rho_X\equiv1+\frac{1}{4\tau}(\mu_X^2-1)\,.
 \label{eq:rhox}
\end{equation}
The $\mathcal{D}$ term is a relativistic correction given by:
\begin{widetext}
\begin{eqnarray} {\cal D}(\kappa,\tau,\rho) &=&
\xi_F\left(1-\psi_X^2\right) 
\left[
1+\xi_F\psi_X^2
-\frac{\lambda}{\kappa}\psi_X
\sqrt{\xi_F\left(2+\xi_F\psi_X^2\right)}
+\frac{\tau}{3\kappa^2}
  \xi_F\left(1-\psi_X^2\right) 
\right]\,.
\nonumber\\
\label{eq:cald}
\end{eqnarray}
\end{widetext}
The $\mp$ sign associated to the $W_3$ function in Eq.~(\ref{eq:Umunu}) refers to neutrino and antineutrino, respectively. Note that for electromagnetic interactions, the $W_3$ term vanishes. More details about the SuSA inelastic formalism can be found in~\cite{megias_charged-current_2017, barbaro_inelastic_2004}.

The integration limits in \eqref{inelastic_response} are given by 
the appropriate 
kinematical restrictions and considering that $W_{X}$ should be above the pion-production threshold. Then the full inelastic spectrum is limited by
\begin{align}
    \mu^{min}_{X}= 1 + \frac{m_{\pi}}{m_{N}}, \\
    \mu^{max}_{X}= 1 + 2\lambda - \frac{E_{shift}}{m_{N}},
\label{limits-inelastic}
\end{align}
where $E_{shift}$ is related to the energy necessary to extract one nucleon from the nucleus . 
The function $f^{model}$ is referred in this approach to the SuSAv2-inelastic scaling function, which exhibits the same functional form of the SuSAv2-QE one but depends on a different scaling variable and is weighted by the invariant mass. The inelastic scaling variable $\psi_X$ is defined as 
\begin{equation}
    \psi_{X} \equiv\frac{1}{\sqrt{\xi_{F}}} \frac{\lambda - \tau\rho_{X}}{\sqrt{(1 + \lambda \rho_{X})\tau + \kappa\sqrt{\tau(\tau\rho^{2}_{X} + 1)}}},
    \label{psix}
\end{equation}
being $\rho_{X}$ given in Eq.~\eqref{eq:rhox}.

Within this formalism, we can also replace the SuSAv2 inelastic scaling function in Eq.~(\ref{inelastic_response}) by a generic one, $f^{model}$, so that other nuclear models where a scaling function can be obtained, such as the RFG, are also applicable.  Here we make use of the SuSAv2 scaling function that takes into account RPWIA and RMF ingredients, being less simplistic than the RFG approach. 

As shown in Eq.~(\ref{limits-inelastic}), the inelastic nuclear response accommodates the whole inelastic spectrum: $\Delta$ resonance, shallow and deep inelastic scattering. However, the integration limits can be modified to exclude some particular contributions. We explore this possibility by excluding the $\Delta$-resonance region from the SuSAv2-inelastic model, so that this approach can be combined with other models for the $\Delta$ production, for example  the SuSAv2-$\Delta$ model (Section~\ref{susav2delta}) or the RMF-1$\pi$ model~\cite{Gonzalez-Jimenez19}, without overlapping. In particular, this is carried out by selecting as the lower limit of integration a value above the $\Delta$ invariant mass (after the $\Delta$ peak, when the $\Delta$ contribution starts to decline), $\mu^{min}_{X} > \mu_{\Delta}$ ($\mu_{\Delta} + 1.5\%\mu_{\Delta}$), in such a way that only nucleon resonances heavier than the $\Delta$ and DIS contribute to the cross section. 
In Eq.~(\ref{cs-only-delta1}), the HR+DIS cross section refers to this approach, {\it i.e.,} the inelastic responses have been evaluated by performing the integral with the lower integration limit being above the $\Delta$ 
invariant mass. 
In the results section (\ref{Results}), we compare the predictions provided by the full SuSAv2-inelastic model for electrons with the ones corresponding to the SuSAv2-$\Delta$ model together with the SuSAv2-DIS approach, i.e., the SuSAv2-inelastic model excluding the $\Delta$ resonance contribution.\footnote{SuSAv2-DIS 
indicates the SuSAv2 treatment of HR+DIS contributions.}

\subsection{Extension of the SuSAv2-DIS model to weak interactions}

The single-nucleon hadronic responses for electromagnetic interactions depend on two inelastic structure functions $W_{1}$ and $W_{2}$ \cite{bjorken_asymptotic_1969} that can be written in a dimensionless form
\begin{align}
  F_{1}=m_{N}W_{1}, \\
  F_{2}=\nu W_{2}, 
\end{align}
where $\nu$ is a Lorentz invariant
\footnote{$\nu\equiv\frac{H\cdot Q}{m_{N}}$, where $H^\mu$ is the 4-momentum of the on-shell nucleon and   $Q^\mu$ the transferred 4-momentum.} coinciding with the transferred energy $\omega$ in the laboratory frame. 
In the deep inelastic regime the two structure functions $F_1$ and $F_2$ (likewise $W_1$ and $W_2$) are linked by the Callan-Gross relation~\cite{PhysRevLett.22.156} $F_{2}=2xF_{1}$, where $x=1/\rho_{X}$ is the Bjorken scaling variable. 

The description of the deep-inelastic regime for weak interactions implies the knowledge of an additional structure function, $F^{\nu}_3 (W^{\nu}_3)$, related to the parity violating contribution associated to the vector-axial interference. 
An accurate determination of this function is hard to achieve from neutrino experiments as well as from parity-violating electron scattering~\cite{doi:10.1146/annurev.nucl.51.101701.132312}  due to the large uncertainties associated to the cross section measurements. However, 
some relationships among the electromagnetic and weak structure functions and between $F^{\nu}_2$ and $F^{\nu}_3$~\cite{bodek_fermi-motion_1981,gluck_dynamical_1998} can be established within the quark-parton model. This is based on the assumption that the corresponding structure functions $W_i$ can be written in terms of quark $Q$ and antiquark $\bar Q$ distributions \cite{bodek_fermi-motion_1981}. This approximation is valid in moderate-large $x$-values, where we neglect strange and charm quarks~\cite{athar_neutrino_2020, bodek_axial_2013, PhysRevC.99.035207}:
\begin{equation}
    F^{\nu N}_2=\nu W^{\nu}_2=Q+\bar Q = x\biggl(u(x) + d(x) + \bar{u}(x) + \bar{d}(x) \biggr), 
    \label{eq:F2}
\end{equation}
 \begin{equation}\label{f3eq}
    xF^{\nu N}_3=x\nu W^{\nu}_3=Q-\bar Q= x\biggl(u(x) + d(x) - \bar{u}(x) - \bar{d}(x) \biggr) , 
\end{equation}
where $u(\bar u)$ and $d(\bar d)$  are the  parton distribution functions (PDFs) 
for the up and down quarks (antiquarks), respectively.

For electron scattering, the isoscalar $F_2$ structure function of the nucleon, defined as the average of the proton and neutron structure functions, is given (at leading order in $\alpha_s$ and for two flavors) by 
\begin{widetext}
\begin{equation}
\label{eq:F2em}
F^{eN}_{2}=\frac{1}{2}\left(F^{ep}_{2}+F^{en}_{2}\right)=\frac{5x}{18}\biggl(u(x) + \bar{u}(x) + d(x) + \bar{d}(x)\biggr). 
\end{equation}
\end{widetext}
The quark distributions are defined to be those in the proton and the factor 5/18 arises from the squares of the quark charges.

  In this region, the weak and electromagnetic $F_2$ structure functions, Eq.~\eqref{eq:F2em} and Eq.~\eqref{eq:F2}, approximately satisfy
\begin{equation}
F^{\nu N}_{2} \approx \frac{18}{5} F^{eN}_{2}.
\end{equation}
Note that this relation is deduced within a quark-parton model from the quark distributions at moderate-large $x$-values,  therefore it is applicable in the regime of very high inelasticity, but fails in the resonance region. 

Thus, under this assumption, which has been tested with experimental results~\cite{HAIDER2015138,Kimprl,Seligman,GRM99}, one can readily obtain the weak inelastic single-nucleon structure functions which are implicit in the terms $G_K^{inel}$ of Eq.~(\ref{inelastic_response}). In this work we describe the structure functions in two different ways. First, we make use of empirical fits of the inelastic electron-proton and electron-deuteron cross sections together with a phenomenological antiquark distribution to extrapolate accurate electromagnetic fits to the neutrino case. The fits employed are Bodek-Ritchie (BR) \cite{bodek_fermi-motion_1981,PhysRevD.24.1400,PhysRevD.20.1471,PhysRevD.12.1884} and Bosted-Christy (BC) \cite{PhysRevC.81.055213,PhysRevC.77.065206}. The BR parameterization fits the SLAC data published in~\cite{PhysRevD.20.1471}, covering a $Q^2$-range from 0.1 to 30 GeV$^2$. The BC fit is constrained by the high precision longitudinal and transverse (L/T) separated cross section measurements from JLab Hall C~\cite{liangjlab} in the kinematic range of four-momentum transfer $0\leq Q^2 \leq 8$ GeV$^2$ and final state invariant mass $1.1 < W_X < 3.1$ GeV, thus going roughly from the pion production region to the highly-inelastic region. 

The second option we explore is to define the inelastic structure functions in terms of the Parton Distribution Function (PDF) model. In particular, we make use of the Gl\"uck-Reya-Vogt  GRV98 model \cite{gluck_dynamical_1998}. In this case, the structure functions are extracted from deep inelastic and other hard scattering processes  at high energies. As a consequence, this parameterization works better at very high values of $Q^{2}$  and $W\gtrsim$3 GeV.

\begin{figure*}[!htbp]
    \centering
    \includegraphics[width=0.856\textwidth]{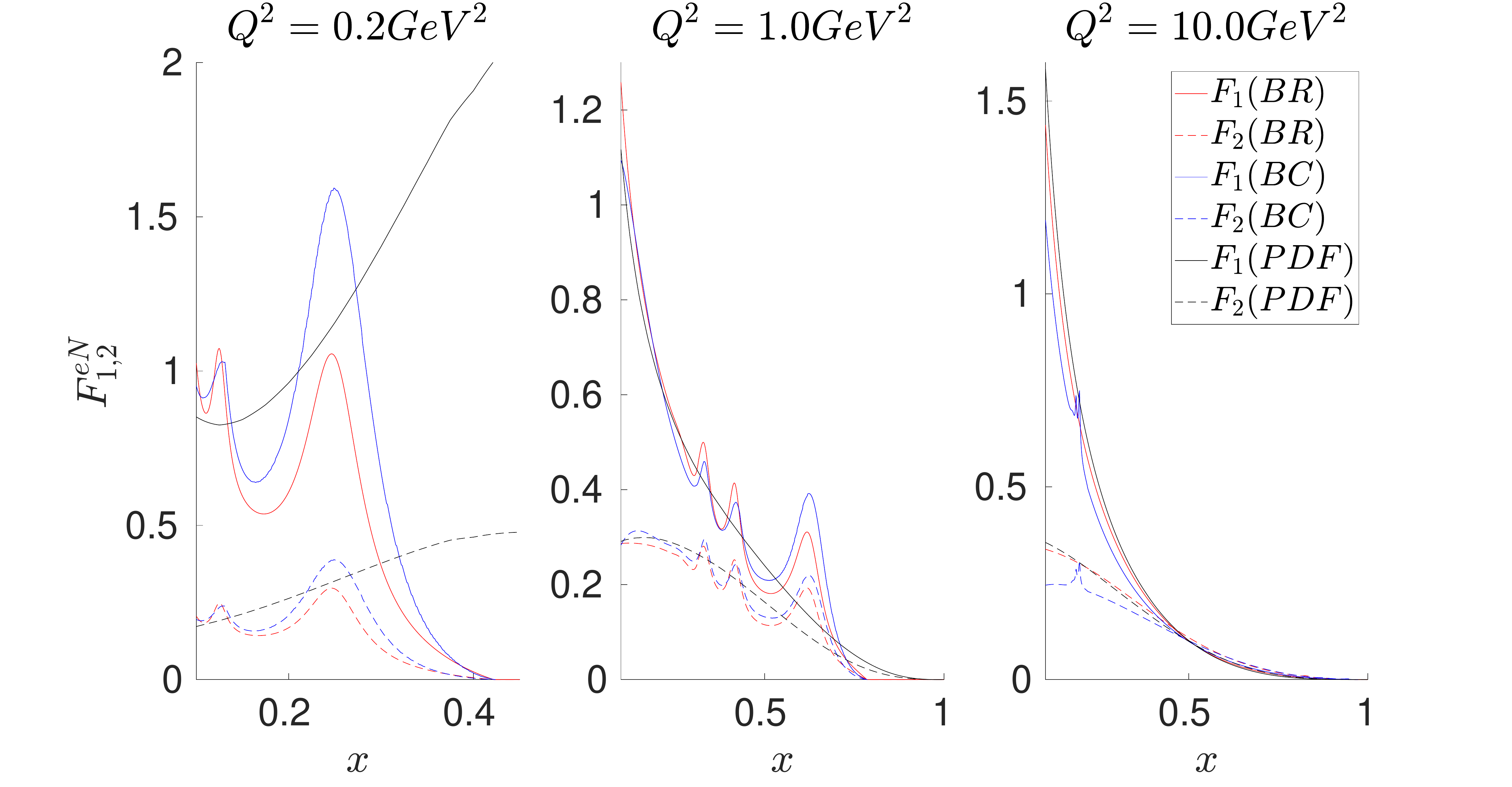}
    \caption{Electromagnetic inelastic nucleon structure functions $F_{1}$ and $F_{2}$ at $Q^{2}=0.2$ GeV$^{2}$ (left), $1.0$ GeV$^{2} $ (center) and $10$ GeV$^{2}$ (right) versus the Bjorken scaling variable $x$ .}  
    \label{F12}
\end{figure*}

In fig.~\ref{F12} we present the results for the two inelastic electromagnetic structure functions $F_1^{eN}$ and $F_2^{eN}$ provided by the three parameterizations discussed above, denoted as BR, BC and PDF. As observed, both BR and BC show the structure of the resonances at lower values of $Q^{2}$, whereas PDF does not, as should be expected. This is consistent with the limitations of the PDF approach at low kinematics. On the contrary, at higher kinematics (panel on the right), the resonance structure is lost and the three models lead to rather similar results although the implementation of the BC fit in the SuSAv2-inelastic model has shown better agreement with ($e,e'$) data than the BR one~\cite{Megias_2016,Amaro_2020}.  
 By combining Eq.~\eqref{eq:F2} and Eq.~\eqref{f3eq}, the additional structure function in weak interactions, i.e. $F_{3}(W_{3})$, can be written as
  \begin{equation}
 xF_{3}^{\nu N}= F^{\nu N}_{2}  - 2\bar{Q}(x)\\, 
 \label{xf3nu}
 \end{equation}
where the antiquark distribution, $\bar{Q}(x)$, can be defined in terms of PDF. However, note that this approach is of limited relevance for current neutrino oscillation experiments where the region of intermediate energies contributes significantly.  Another option is to define the antiquark distribution in terms of the empirical fits of electron scattering from BR parameterization \cite{bodek_fermi-motion_1981}, which works at low-intermediate kinematics. In the analysis that follows we make use of this second option, unless otherwise stated.

These methods of extending electromagnetic inelastic structure functions cannot properly handle the $\Delta$-region in the case of neutrinos.  


\begin{figure*}[!htbp]
    \centering
    \includegraphics[width=0.856\textwidth]{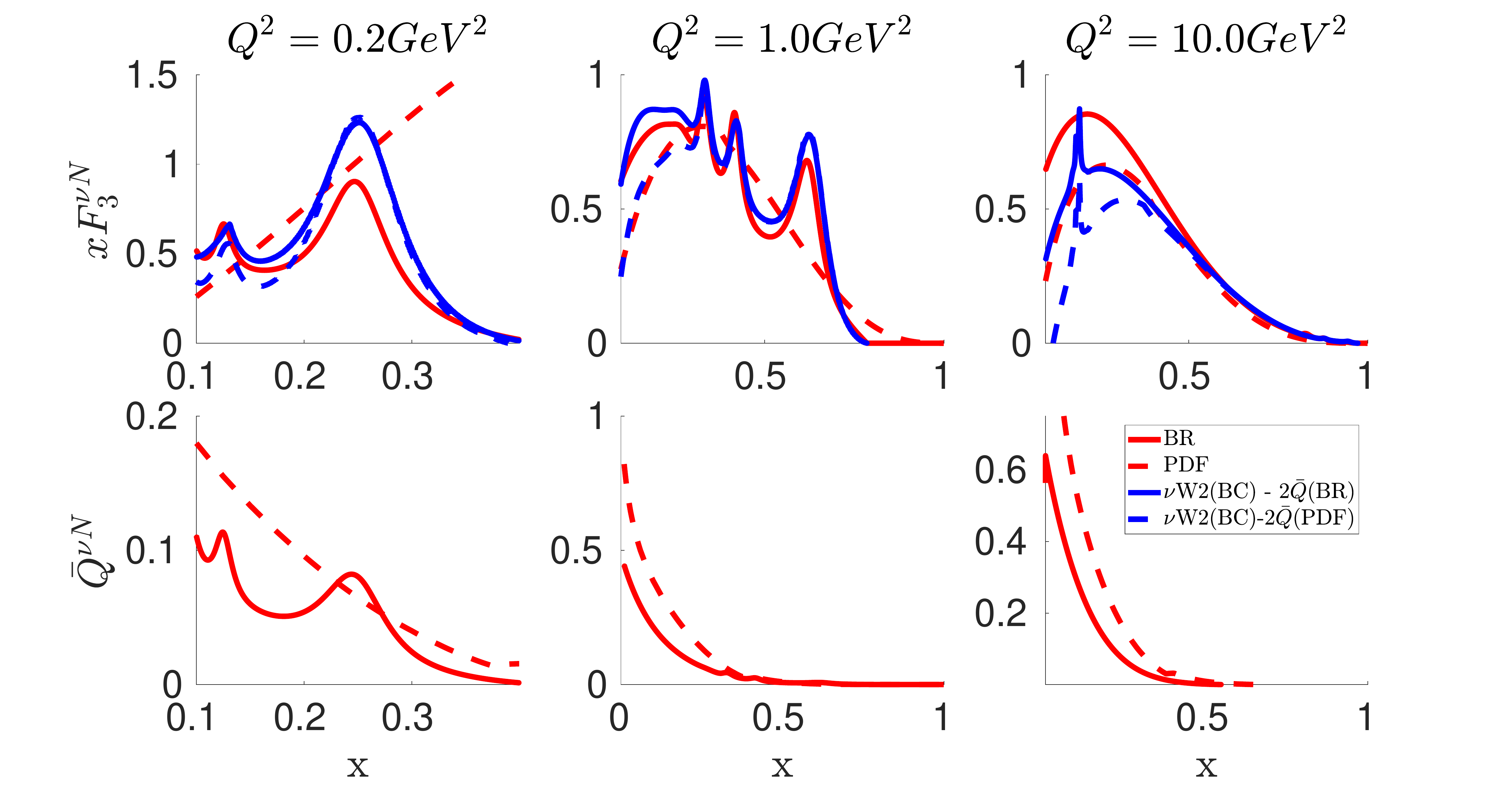}
    \caption{Weak inelastic nucleon structure $xF_{3}$ at $Q^{2}=0.2$ GeV$^{2}$ (left), $1.0$ GeV$^{2}$ (center) and $10$ GeV$^{2}$ (right).
      }
    \label{F3}
\end{figure*} 

The results for $xF_3^{\nu N}$ (top panels) and  $\bar{Q}^{\nu N}$ (bottom) are shown in fig.~\ref{F3}. In the case of $\bar{Q}^{\nu N}$ we compare the predictions corresponding to Bodek-Ritchie (solid red line) and GRV98 (dashed red line). As noticed, a significant discrepancy is observed, particularly at the lowest value of $Q^2$ (left-bottom) where BR shows the structure associated to the nucleon resonances whereas GRV98 does not. On the contrary, at higher $Q^2$ the BR  and GRV98 curves behave similarly although the latter is significantly larger. These results are consistent with the PDF model that only works at high values of the transferred four-momentum. 
Similar comments also apply to results shown for $xF_3^{\nu N}$, although here the BR result exceeds the GRV98 one in the maximum at high kinematics. For completeness, we also present the results for $xF_3^{\nu N}$ evaluated using eq.~(\ref{xf3nu}), but with parameterization and the antiquark distribution, $\overline{Q}(x)$, calculated with BR (blue solid line) and GRV98 (blue dashed). In this case the nucleon resonance structure at low-intermediate $Q^2$ is clearly shown with similar results for the two prescriptions. Moreover, although differences are observed with the previous calculations (red lines), the general behavior of the results follows a similar trend for all the models.

\section{Results \label{Results}}


The description of inclusive electron and/or neutrino scattering processes requires to take into account the contribution of different reaction channels. These are summarized in Table \ref{TableContribution} together with the theoretical models used to describe them. Also included are the corresponding abbreviations written in the legends of the subsequent graphs. At transferred energy $\omega\simeq Q^2/2m_N$, the dominant process is quasielastic (QE) scattering that is described with the SuSAv2 superscaling model introduced in the previous section.
As the value of $\omega$ increases,  2p2h states can be excited via Meson Exchange Currents (MEC). This is modelled within the framework of the Relativistic Fermi Gas (RFG-MEC). 
At higher energy transfer a pion can be emitted through the excitation of the $\Delta$ resonance $(\Delta)$. This process is described by 
the SuSAv2-$\Delta$ model  introduced in Section~\ref{susav2delta}. Finally, the deep inelastic scattering (DIS) region that occurs at the highest energies is accounted for by the SuSAv2 inelastic model (Section \ref{SuSav2-inel}). Notice that by DIS we denote not only the process in which the probe interacts with the partons, but also the region where nucleon resonances heavier than the $\Delta$ are excited. 
We also present results where the SuSAv2 inelastic model is used to describe the full inelastic ($e,e'$) spectrum, i.e., including also the contribution of the $\Delta$ resonance. 
  
\begin{table*}
\begin{tabular}{|c|c|c|} 
\hline
\textbf{Abbreviation} & \textbf{Contribution} & \textbf{Model} \\
 \hline
 QE & Quasielastic & \parbox[t]{2cm}{SuSAv2 \\ superscaling} \\ 
 \hline
MEC &   2p2h excitations & RFG-MEC  \\ 
 \hline
 $\Delta$ & $\Delta$ resonance & SuSAv2-$\Delta$  \\
 \hline
DIS& \parbox[t]{5cm}{ Higher resonances \\  and deep inelastic }   &SuSav2 inelastic  \\ 
 \hline
Full Inelastic & Whole inelastic spectrum  & SuSAv2 inelastic   \\ 
 \hline
\end{tabular}
\caption{\label{TableContribution} Channels that contribute to the reaction mechanism with the notation followed in the text and the model used to evaluate the cross section. The chosen inelastic structure functions are given by the Bodek-Ritchie (BR), Bosted-Christy (BC) or Parton Distribution Functions (PDF) prescriptions.}
\end{table*}

\subsection{Electron Scattering \label{electron-result}}


As a first step we test the models presented in the previous section versus inclusive electron scattering data, $(e,e^{\prime})$. Here we show
results for some representative choices of kinematics, similar to those involved in neutrino scattering processes, and different nuclei.



  \begin{figure*}[!htbp]
    \centering
    \includegraphics[width=\textwidth]{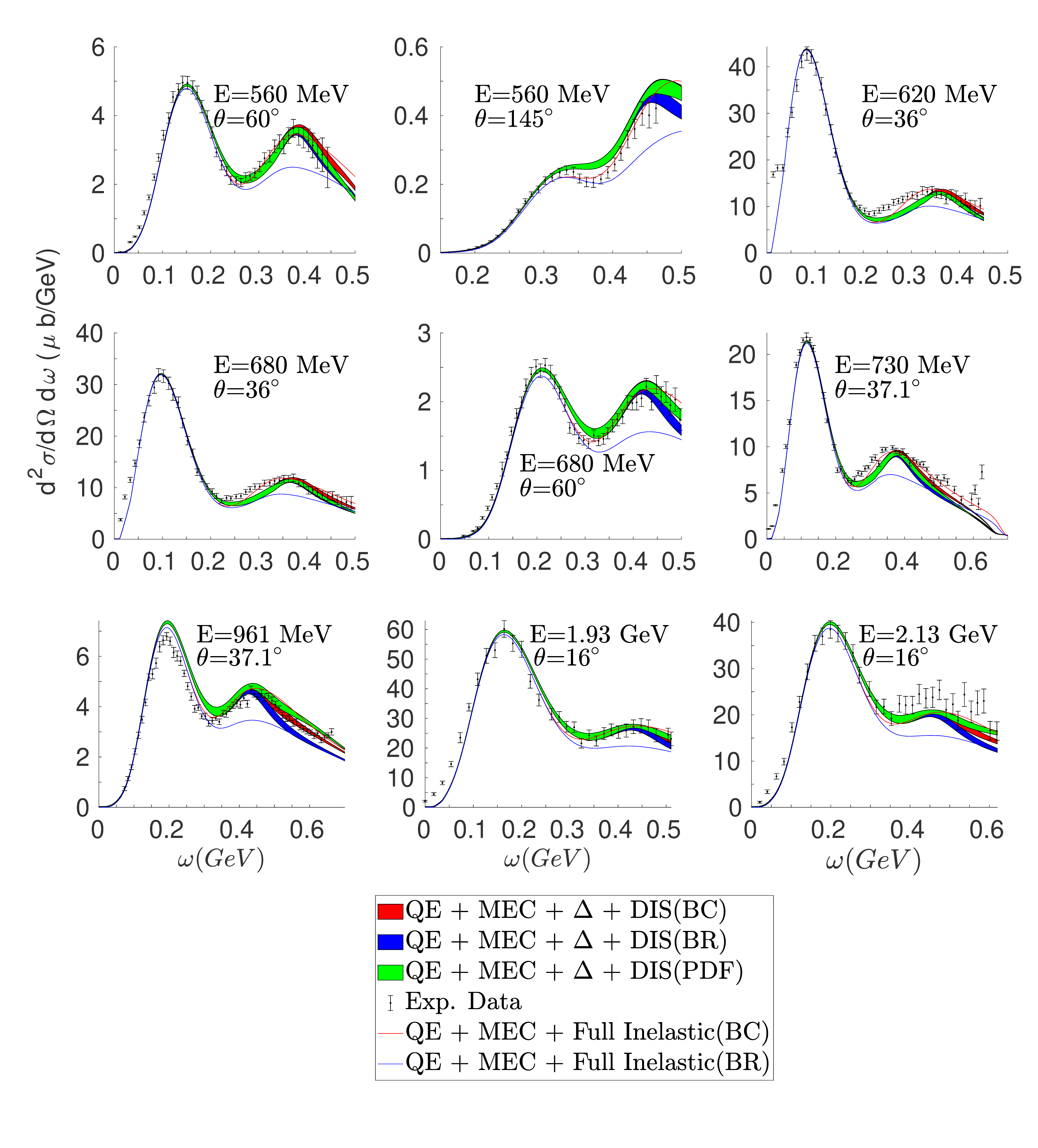}
    \caption{\label{C12}  Double-differential inclusive cross section for e-$^{12}C$ scattering at given beam energies and scattering angles (labeled in the panels). It is displayed in function of the transferred energy. The notation in the legend refers to Table \ref{TableContribution}. Data from \cite{benhar_archive_2006}. 
    }
    \label{jac1}
\end{figure*}

\begin{figure*}[!htbp]
	\centering
	\includegraphics[width=\textwidth]{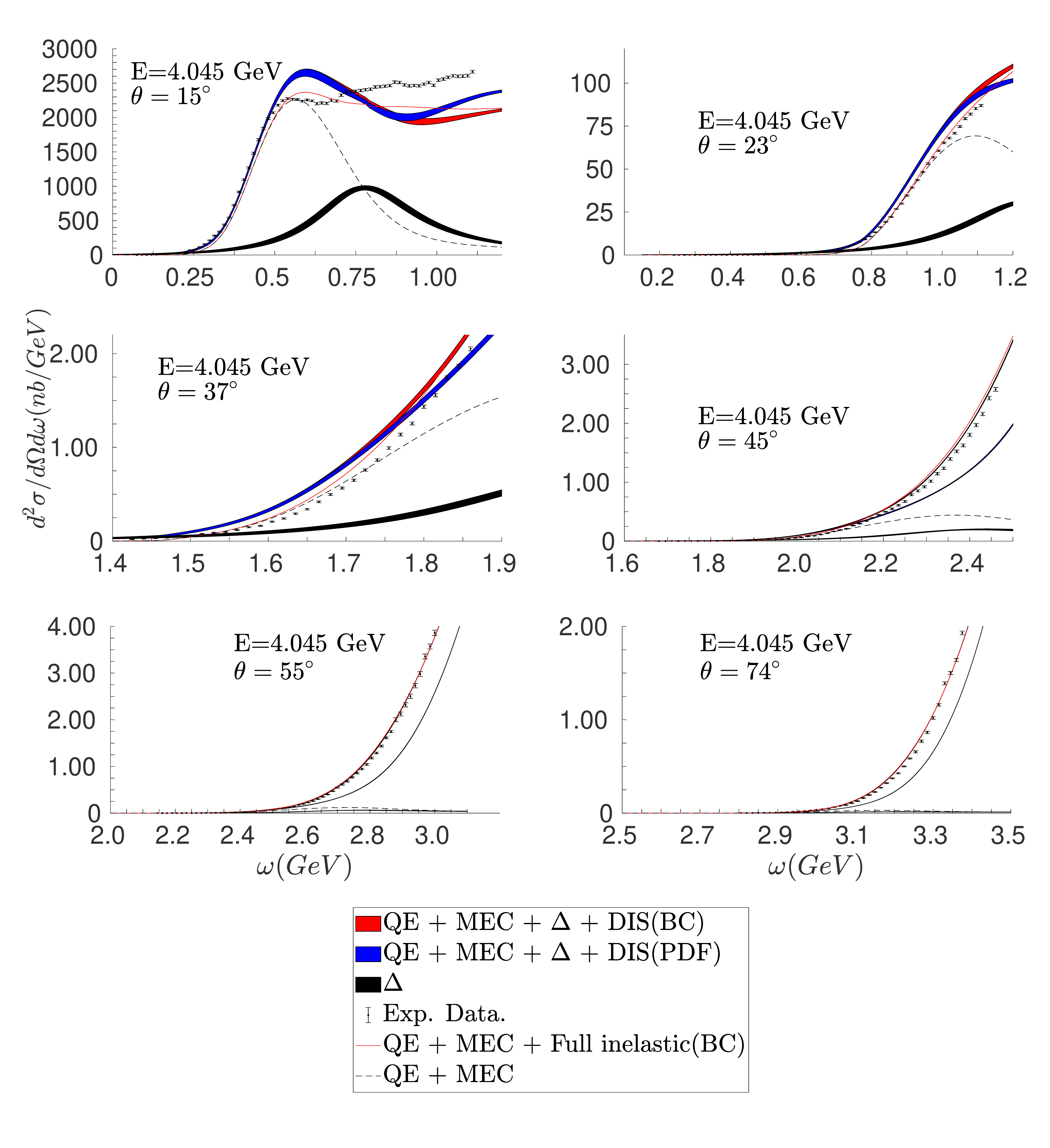}
	\caption{\label{JLab12C} Same as fig.~\ref{jac1}, except now showing the separate contributions for QE+MEC and $\Delta$. Data from \cite{benhar_archive_2006}. 
	}
\end{figure*}

In fig.~\ref{C12} we show the electron-carbon cross section versus the energy transfer $\omega$ for electron beam energies ranging from 560 to 2130 MeV and different scattering angles. At these kinematics the dominant processes are QE scattering and $\Delta$ production. This is clearly illustrated in the figure by the two broad peaks. The dip region between the two peaks is filled by the 2p2h contribution. 
 While the QE and MEC contributions are calculated using the SuSAv2  and RFG models, respectively, we explore different options for the treatment of the inelastic processes, corresponding to the different curves in each plot. We observe that the total result obtained using $\Delta$+DIS provides a reasonably good description of the data, independently of the parameterization used for the elementary structure functions - BR, BC or PDF.
 On the contrary, the full inelastic model gives good agreement with the data only in the case of the BC parameterization, while BR underestimates the data in the region of the $\Delta$ peak. This can be explained by noticing that at these kinematics the value of $Q^2$  is below 1 GeV. 
As shown in fig. \ref{F12}, in this region the results of the two prescriptions are significantly different. Note also that the kinematics considered are not appropriate to use the full inelastic model with PDF \cite{gluck_dynamical_1998}.

Fig.~\ref{JLab12C} shows the comparison between our predictions and data at much higher electron energy, E$\sim$4 GeV. In this case the 
$\Delta$ and DIS channels give a very sizeable contribution, becoming dominant as the scattering angle increases. 

As already mentioned, we use two different methods to get the inelastic cross section. One consists in using the superscaling function folded with the inelastic BC structure functions to describe the full inelastic spectrum, the other combines the scaling function $f^\Delta$ in the $\Delta$ resonance region (``SuSAv2-$\Delta$'') and the DIS model  (``SuSAv2 inelastic") beyond (with BC and PDF). 
As noticed, the prediction provided by the full inelastic model (QE+MEC+Full inelastic) fits nicely the data in most of the kinematics situations, although in some cases, {\it i.e.,} intermediate values of the scattering angle, 37 and 45 degrees, the theoretical models tend to overpredict data. The combination of SuSAv2-$\Delta$ and SuSAv2 inelastic introduces a band due to the uncertainty of the statistical analysis in the determination of the $\Delta$ scaling function. Here the results,  presented by the red band corresponding to BC  clearly overestimate the data at the smaller angles, whereas the agreement improves significantly for larger values, i.e., 55 and 74 degrees. In the case of PDF (blue band), the predictions are below the other two parameterizations underestimating the data for  45, 55 and 74 degrees and it shows a similar behavior to BC at lower angles.  

 \begin{figure*}[!htbp]
    \centering
    \includegraphics[width=\textwidth]{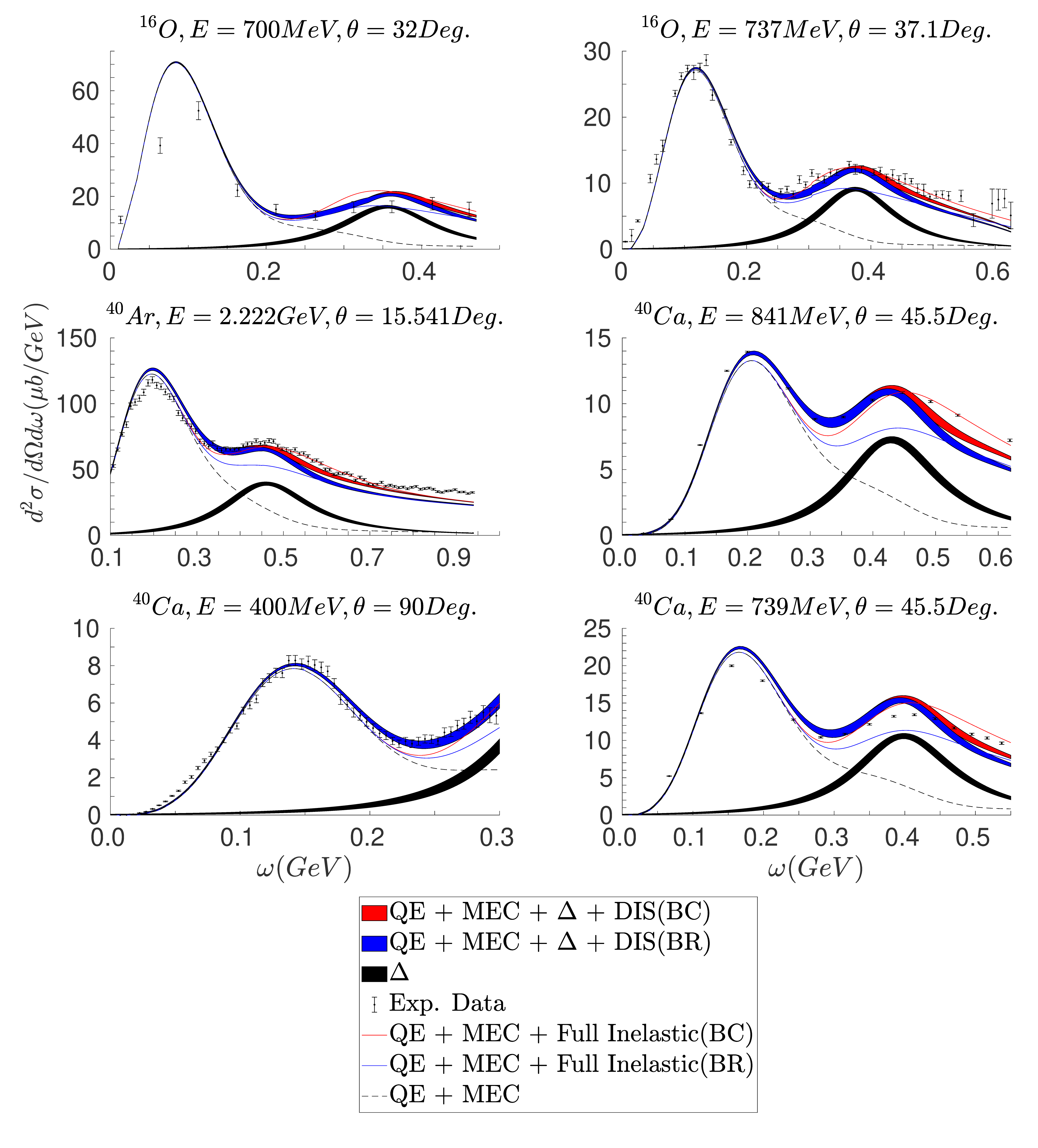}
    \caption{\label{other-nuclei} Same as fig. ~\ref{jac1}, except now for oxygen (top panels), argon (middle) and calcium (bottom) and different kinematics. Data from \cite{benhar_archive_2006}.}
\end{figure*} 


In fig.~\ref{other-nuclei} we extend our analysis to other different nuclear systems: oxygen, argon and calcium. In each case we compare the data with the predictions of our models, QE+MEC+$\Delta$+DIS and QE+MEC+Full inelastic, with the two parameterizations for the inelastic single-nucleon structure functions: BR and BC. As observed, although the general behaviour of data is successfully described by the models, significant discrepancies between them are shown. In the case of the QE+MEC+Full inelastic, BC and BR prescriptions lead to very different results in the region where the $\Delta$ gets its maximum, being BC much larger and closer to the data. On the contrary, the discrepancy between the two models - BC and BR- associated to the QE+MEC+$\Delta$+DIS description is significantly smaller. In some cases the two bands overlap providing in general a good description of the data. In fact, these predictions agree with the ones corresponding to the QE+MEC+Full inelastic using BC. 
 


In general, the QE + MEC + Full inelastic model with BC describes well the electron scattering data. However, this model cannot be properly extended to neutrinos as stated in section \ref{SuSav2-inel}. In the case of QE + MEC + $\Delta$ + DIS, the three models considered for the DIS contribution provide predictions that agree with electron scattering data. This gives us confidence in their applicability to the analysis of neutrino-nucleus scattering reaction. This analysis is presented in the next subsection.

 

\subsection{Neutrino Scattering\label{neutrino-result}}

 

In this section we apply the models summarized in Table \ref{TableContribution} and tested against electron scattering  to the analysis of inclusive neutrino-nucleus scattering processes corresponding to different experiments: T2K, MINERvA, MicroBooNE and ArgoNEUT. We compare our predictions with the data for a very wide range of kinematical regimes. In spite of the good description provided by the QE+MEC+Full inelastic model for electron scattering data, here we restrict our attention to the use of the QE+MEC+$\Delta$+DIS. The use of the QE+MEC+Full inelastic model applied to neutrino reactions in the full inelastic regime can be questionable as  the approach considered to get the weak inelastic $W_i$ structure functions from the electromagnetic ones relies on the quarks description which is suited for the DIS regime, but fails in the resonance region. 





\subsubsection{T2K \label{T2K_result}}

\begin{figure*}[!htbp]
  \includegraphics[width=\textwidth]{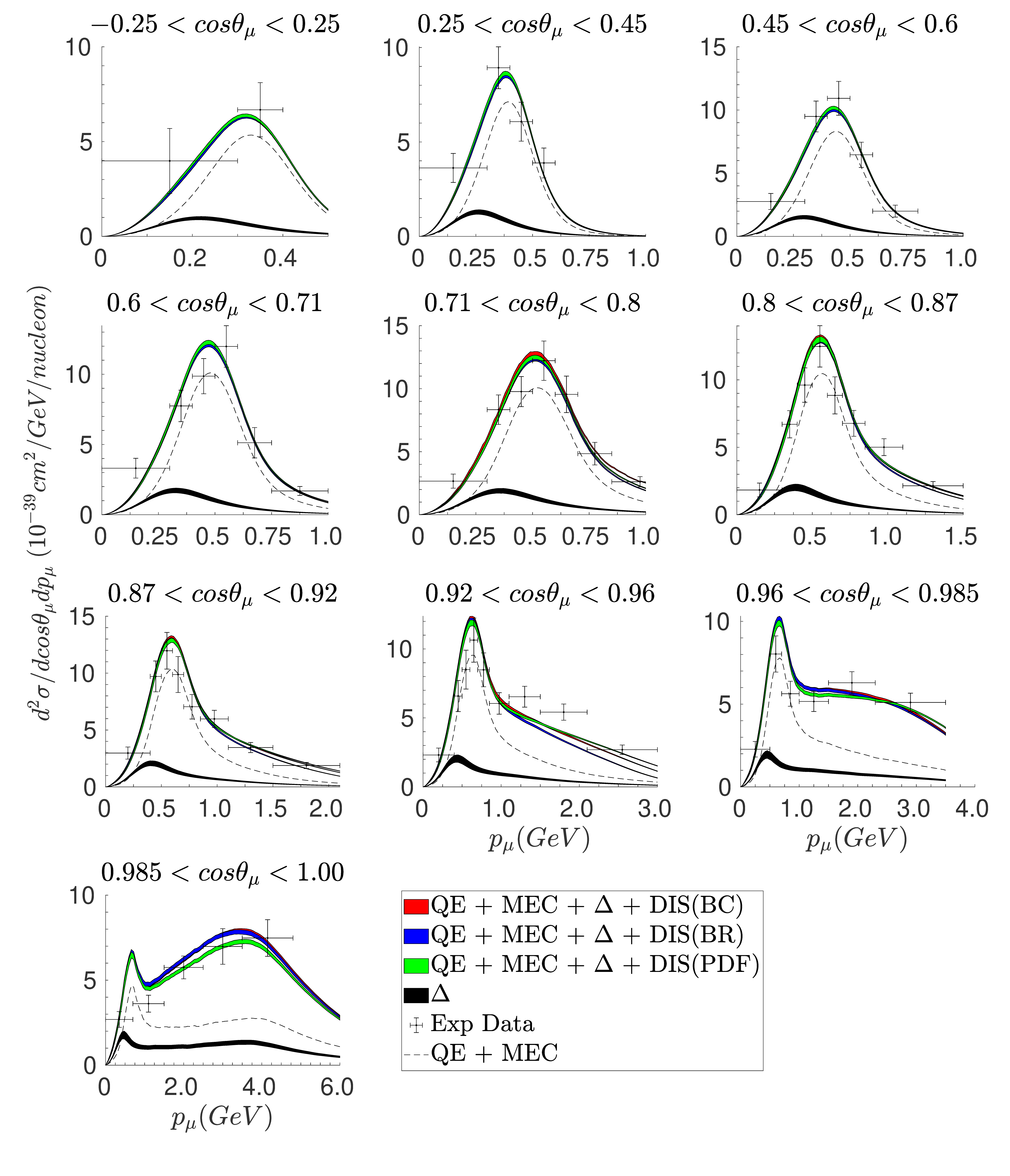}
    \caption{T2K CC inclusive flux-averaged double-differential cross section per target nucleon in bins of the muon scattering angle as function of the muon momentum. Legend as in previous figures (see Table \ref{TableContribution}). Data taken from \cite{Abe_2018}.
    \label{T2K} }
\end{figure*}

In the T2K experiment, the neutrino flux is peaked at 0.6 GeV and the target used in the near detector is carbon \cite{Abe_2018}. 
In fig. \ref{T2K} we show the CC-inclusive $\nu_{\mu}-^{12}$C double-differential cross section per nucleon versus the muon momentum, $p_{\mu}$, for different angular bins, folded with the T2K flux. The different channels that contribute to the cross section are shown separately. As observed, the QE dominates the cross section for values of the scattering angle $\gtrsim$ 30$^\circ$, corresponding to values of muon momentum 
$\lesssim$ 1.5 GeV (panels on the first and second rows).
 As more forward kinematics are explored the muon momentum values allowed by kinematics get larger and the relative contribution of the DIS is more and more important (although not directly displayed in the figure, the significant contribution of this channel is clearly visible by subtracting from the total prediction the contributions of the other channels). This is clearly shown by observing the panels on the lower rows. On the other hand, notice that the contribution of the $\Delta$, although smaller than the QE + MEC one, is clearly visible for all kinematics.

Finally, the results shown by the red, blue and green bands correspond to the sum of all channels using the BC, BR and PDF parameterizations, respectively. As observed, the predictions of the three models are rather similar (only departing for some particular kinematics at forward angles)  
and provide excellent agreement with most of the data. Only at the most forward angles the models tend to overstimate the data in the region of the QE peak at the most forward angles and low $p_\mu$ where the scaling approach may fail. This can be addressed using the RMF model. 

In a recent work by Martini et al.~\cite{Martini:2022ebk} the same data were analyzed using a model based on Random Phase Approximation within the basis of a local Fermi gas calculation. While the results of this study are very similar to the ones presented here for $\cos\theta_\mu\lesssim$0.9, in spite of the different theoretical approach, at very forward angles a better agreement with the data is achieved around the QE peak, signalling that long-range RPA correlations, which are absent in our model,  play an important role at these kinematics. On the other hand at high values of $p_\mu$ we get a better description of the data because the calculation \cite{Martini:2022ebk} does not include inelastic channels beyond one-pion production.

%


\subsubsection{MINERvA \label{Minerva_result}}

In MINERvA experiment the target is hydrocarbon and the neutrino energy flux is peaked at 3.5 GeV \cite{Filkins_2020}. This value is much larger than the one corresponding to T2K, so the contribution of the inelastic channel is expected to be much stronger in MINERvA. Here the data are given in function of the longitudinal and transverse muon momentum that are defined as $p_L=p_\mu \cos\theta_\mu$ and $p_T=p_\mu \sin\theta_\mu$, with $\theta_\mu$ the muon scattering angle. According to the MINERvA acceptance, the muon scattering angle is limited to 
$\theta_{\mu}<$20$^o$ and the muon momentum to 1.5 GeV $<p_{L}<$20 GeV,  $p_{T}<$2.5 GeV. 
 
\begin{figure*}[!hptb]
		 \includegraphics[width=0.85\textwidth]{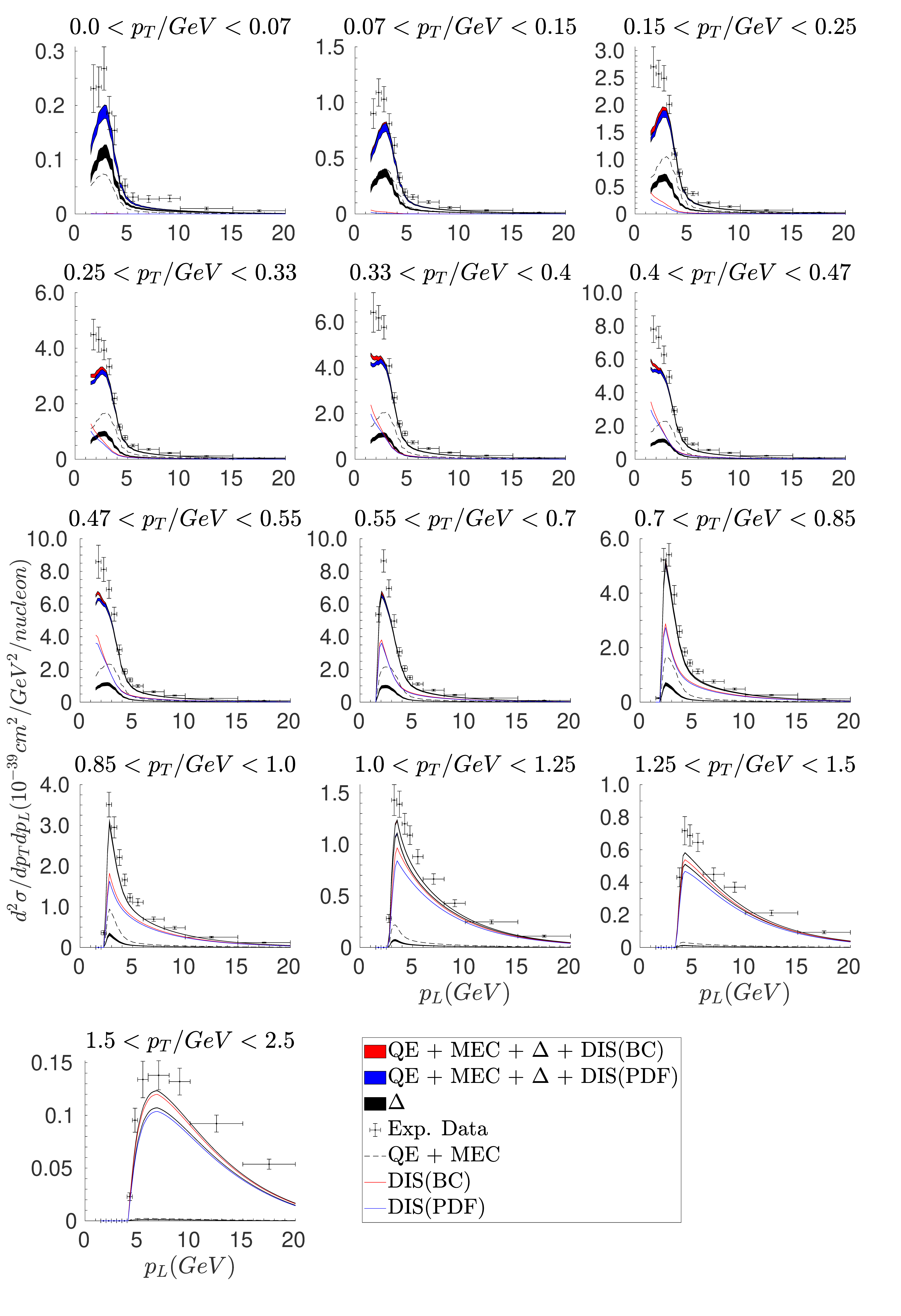}\vspace*{-0.5cm}
		\caption{\label{Minerva_dpl} The CC-inclusive Minerva flux-folded $\nu_{\mu}$-$^{12}$C double differential cross section per nucleon in bins of the muon transverse momentum. The cross section is displayed as a function of the muon longitudinal momentum. Legend referred to Table \ref{TableContribution}. Data taken from \cite{Filkins_2020}.}	 
\end{figure*}

\begin{figure*}[!hptb]
		 \includegraphics[width=\textwidth]{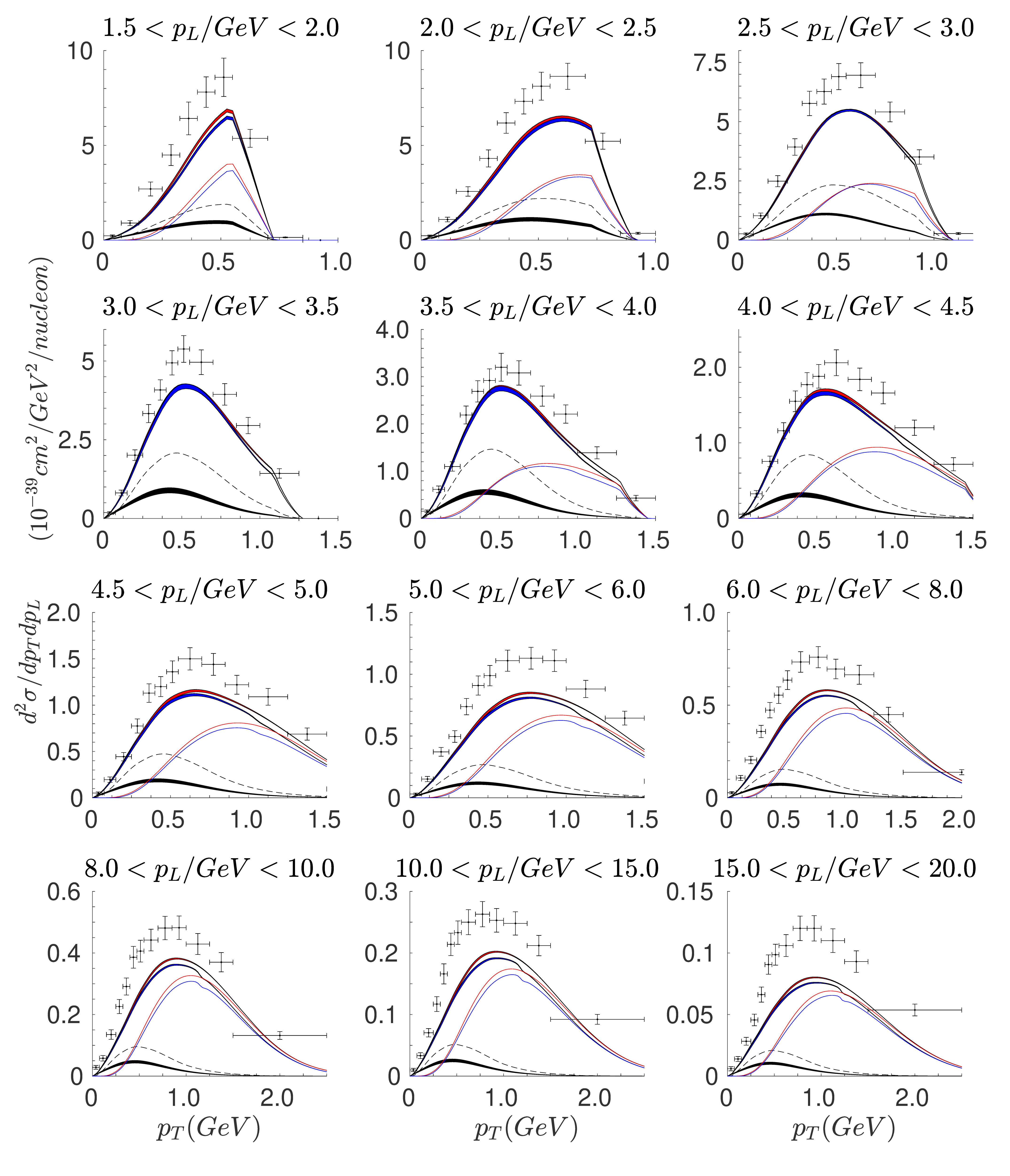}	
	\caption{\label{Minerva_dpt} Same as in fig. ~\ref{Minerva_dpl}, but with the cross section in bins of the muon longitudinal momentum and displayed against the transverse component, $p_T$. Data taken from \cite{Filkins_2020}.}
\end{figure*}

In figs.~\ref{Minerva_dpl} and \ref{Minerva_dpt} we show the CC-inclusive $\nu_{\mu}-^{12}$C double-differential cross section per nucleon versus the longitudinal or transverse momentum, $p_{L},p_{T}$, for different momentum bins, folded with the MINERvA flux. Results are very similar for the two parameterizations, BC and BR, of the inelastic single-nucleon structure functions. For this reason, we only show BC and PDF parameterizations.
As the transverse momentum - and hence the scattering angle - increases the DIS channel becomes more important. This is clearly shown in the results presented in both 
figs.~\ref{Minerva_dpl} and \ref{Minerva_dpt}, and it is also consistent with the behaviour shown in fig.~\ref{JLab12C} for electron scattering. Notice that the theoretical predictions agree with the general shape shown by the data, although a significant discrepancy is observed in the region where the QE dominates (calculations underestimate data by around 25-30$\%$). 


   \begin{figure*}[!hptb]
	\centering
		 \includegraphics[width=1.0\textwidth]{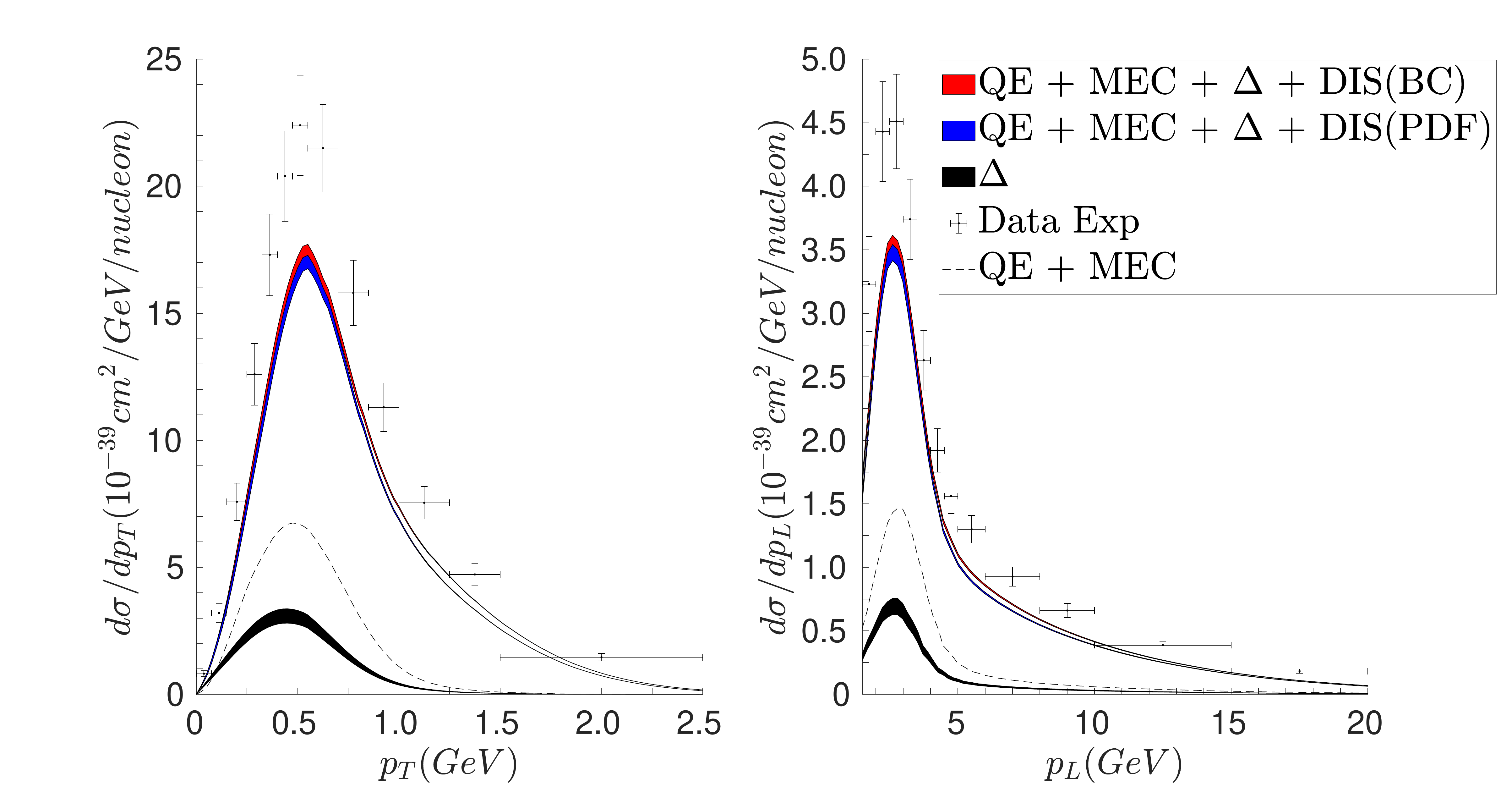}	
	\caption{\label{Minerva_single} The CC-inclusive Minerva flux-folded $\nu_{\mu}$-$^{12}$C single differential cross section per nucleon as function of the muon longitudinal (right) and transverse (left) momentum. Legend as in previous figures (see Table \ref{TableContribution}). Data from \cite{Filkins_2020}.}
\end{figure*}

In fig.~\ref{Minerva_single} we present the CC-inclusive $\nu_{\mu}-^{12}$C single-differential cross section per nucleon folded with the 
MINERvA flux. The panel on the left (right) shows the cross section of the transverse $p_T$ (longitudinal $p_L$) component of the muon momentum. As in the previous case for the double differential cross section, the model QE+MEC+$\Delta$+DIS has been considered with two parameterizations, BC and PDF, for the inelastic nucleon structure functions. The separate contributions of the different channels, QE + MEC and $\Delta$ are shown. It should be pointed out the relevance of the DIS contribution in the whole range of momentum explored. Its contribution in the maximum of the cross section is of the same order of even larger than the QE+MEC response. Furthermore, whereas the DIS maximum is clearly shifted to the right in the left panel (larger values of $p_T$) compared with the QE+MEC and/or $\Delta$, the opposite occurs for $p_L$ (right panel), although here the shift is much less pronounced. On the other hand, the two prescriptions, BC and PDF, lead to minor discrepancies. The difference between both bands are more noticeable in fig. \ref{Minerva_dpl} at higher transverse momenta, where the peak is lower for PDF parameterization.  In other cases, they tend to overlap.

Finally, regarding the comparison with the experiment, we observe that the models reproduce the general shape and behavior of the data, although a significant discrepancy is shown in the maximum of the cross section. Theoretical predictions underestimate data by $\sim 20\%$. This result is consistent with the double differential cross sections presented in figs.~\ref{Minerva_dpl} and \ref{Minerva_dpt}. Comparing with MnvGenie (tuned) from \cite{Filkins_2020}, QE + MEC contributions match with the results portrayed in the paper and it seems to indicate that the model predictions for the DIS and/or $\Delta$ are too small for MINERvA kinematics, while they are doing well for T2K. Although this can be connected with the very different neutrino energies involved in the two experiments and the energy spectrum explored, further studies are needed to clearly establish the validity of the models and their applicability regime.

In previous studies  \cite{Megias_2019}, the prediction of the SuSAv2-MEC has been tested versus antineutrino  CC QE-like MINERvA results. In these cases, the results reproduce well the data without underestimation. Nonetheless, these data do not contain inelastic channels, unlike the ones presented here.



\subsubsection{MicroBooNE \label{Microboone_result}}

\begin{figure*}[!htbp]
    \centering
       \includegraphics[width=\textwidth]{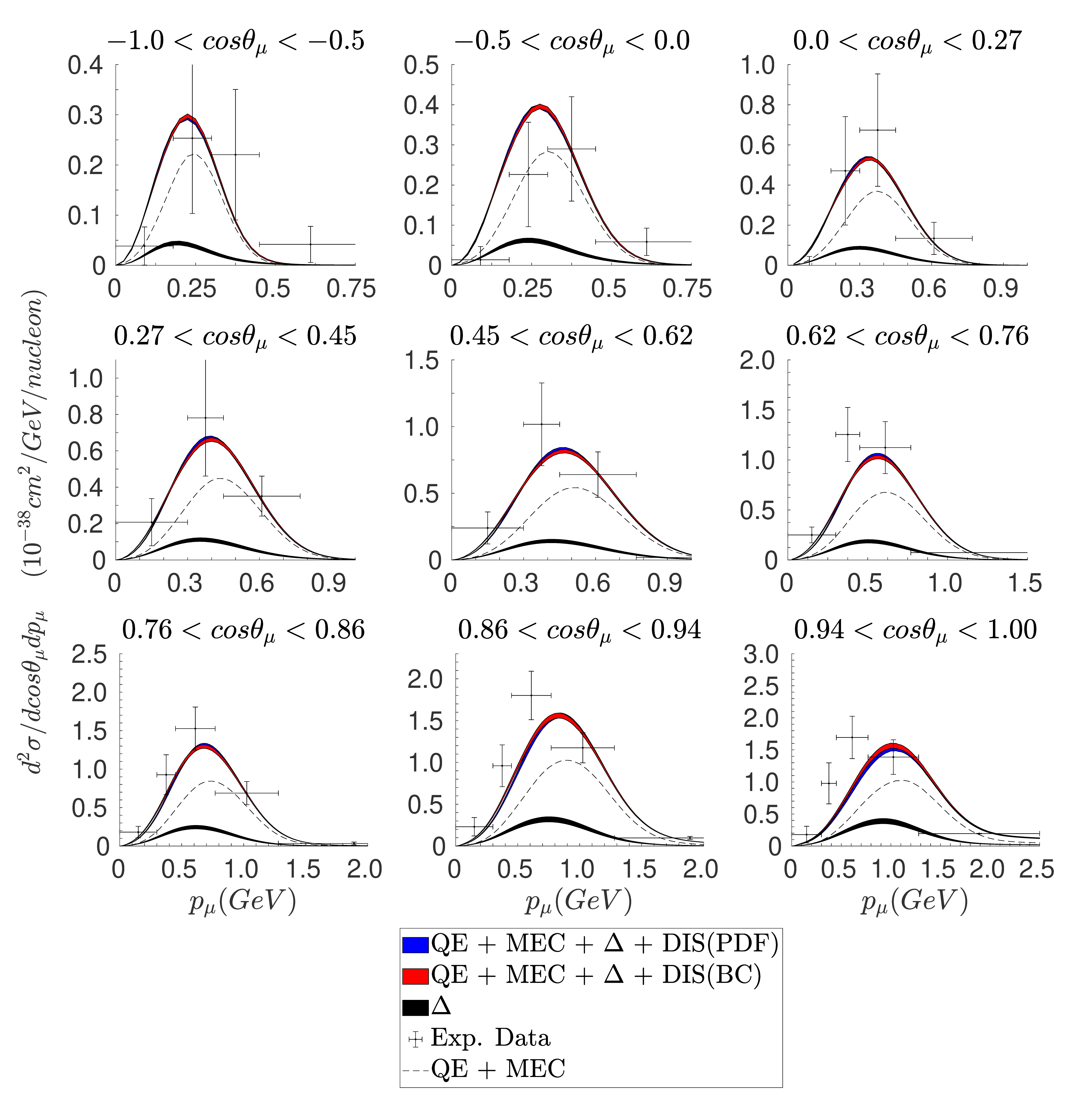}
     \caption{The CC-inclusive MicroBooNE flux-folded $\nu_{\mu}$-$^{40}$Ar double differential cross section per nucleon in bins of the muon  scattering angle as function of the muon momentum. Legend as in previous figures (see Table \ref{TableContribution}). Data taken from \cite{Abratenko_2019}. \label{MicroBOONE}  }
    \end{figure*}

In this experiment, the neutrino beam flux is peaked at $\sim 0.8$ GeV, and the target is liquid argon \cite{Abratenko_2019}. In fig. \ref{MicroBOONE} we show the CC-inclusive $\nu_{\mu}-^{40}$Ar double-differential cross section per nucleon versus the muon momentum, $p_{\mu}$, for different angular bins, folded with the MicroBooNE flux. The same models used for T2K and MINERvA are considered here, and the isolated contributions of the different channels are also displayed. As observed, the discrepancy introduced by the particular description of the inelastic nucleon structure functions, BC or PDF, is negligible, {\it i.e.,} the two color bands overlap for all kinematics. Similar results were observed also for BR.
Concerning the role played by the different channels, it is clearly shown that the QE regime gives the maximum contribution, approximately $55-60\%$ of the total response. The remaining $40-45\%$ strength comes from the inelastic channels that result necessary to explain the data. Concerning the specific role of the $\Delta$ and DIS both produce a similar contribution in most of the cases. Only at the most forward angles the $\Delta$ response is around twice the contribution of DIS. On the opposite, for larger angles, although the global strength of both channels is similar, the shape of the curves differs a little, being the DIS maximum located at smaller values of $p_\mu$. 

Finally, the models provide in general a reasonable description of data. Because of the large error bands it is difficult to draw clear conclusions on the different ingredients involved in the description of the process. However, it is clearly observed that at backward angles the prediction of the models is slightly shifted to lower values of $p_\mu$ compared with data, whereas the reverse occurs at very forward angles. This behavior is consistent with the MonteCarlo analysis for MicroBooNE \cite{Abratenko_2019} as well as with the recent results of Ref.~\cite{Martini:2022ebk}. Moreover, the general agreement between our calculations and data confirms also previous analyses based on simulations with GENIE presented in \cite{Abratenko_2019}.

 
\subsubsection{ArgoNEUT\label{Argoneut_result}}

\begin{figure*}[!htbp]
    \centering\hspace*{-1.25cm}
    \includegraphics[width=\textwidth]{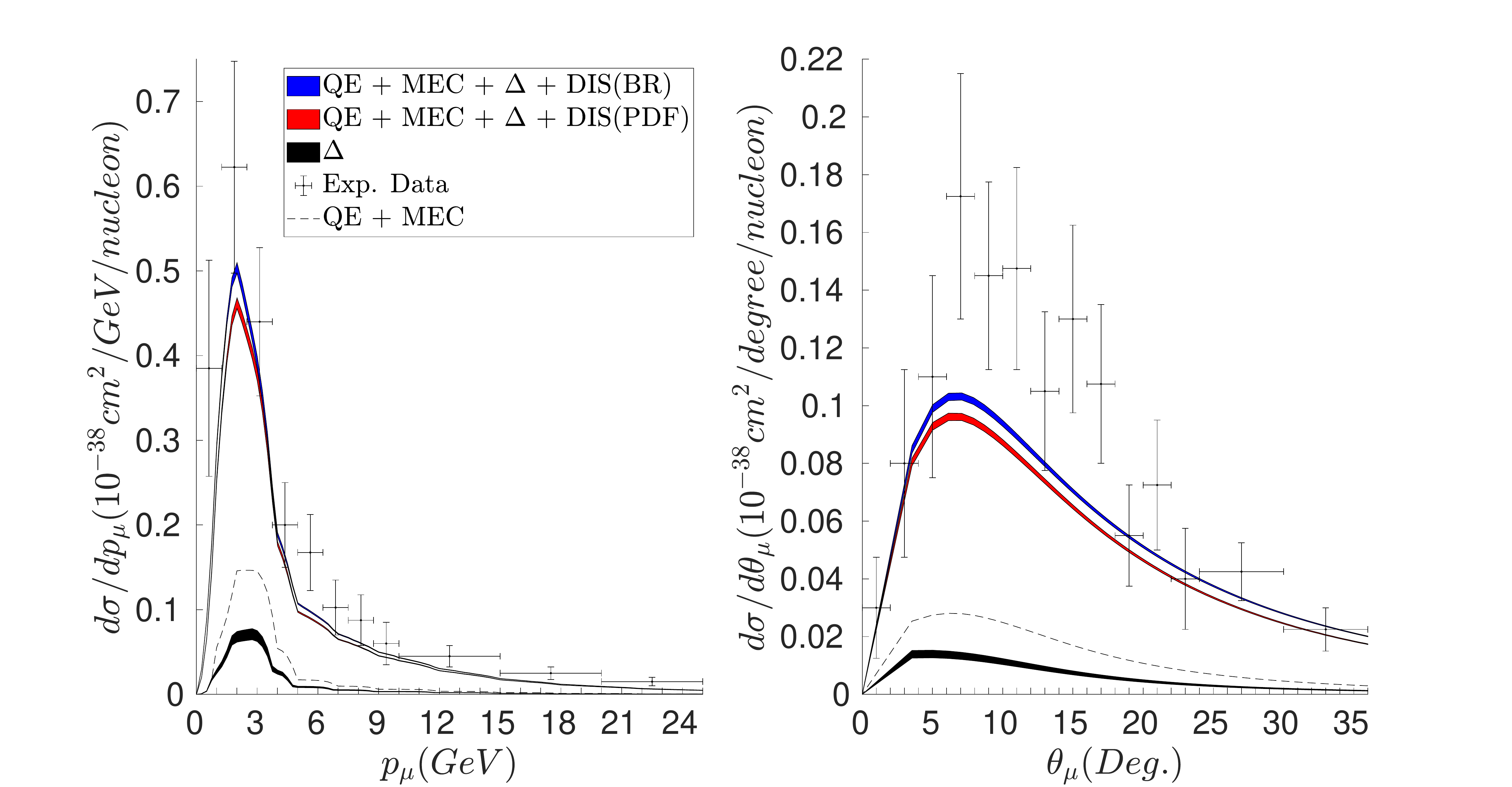}
    \caption{The CC-inclusive ArgoNEUT flux-integrated $\nu_{\mu}$-$^{40}$Ar single differential cross section per argon nucleon, displayed as function of the muon momentum (left panel) or the muon scattering angle (right).  Legend as in previous figures (see Table \ref{TableContribution}). Data taken from \cite{ArgoNeuT:2011bms}. 
\label{ArgoNEUT_old} }
\end{figure*}

\begin{figure*}[!htbp]
    \centering\hspace*{-1.25cm}
    \includegraphics[width=\textwidth]{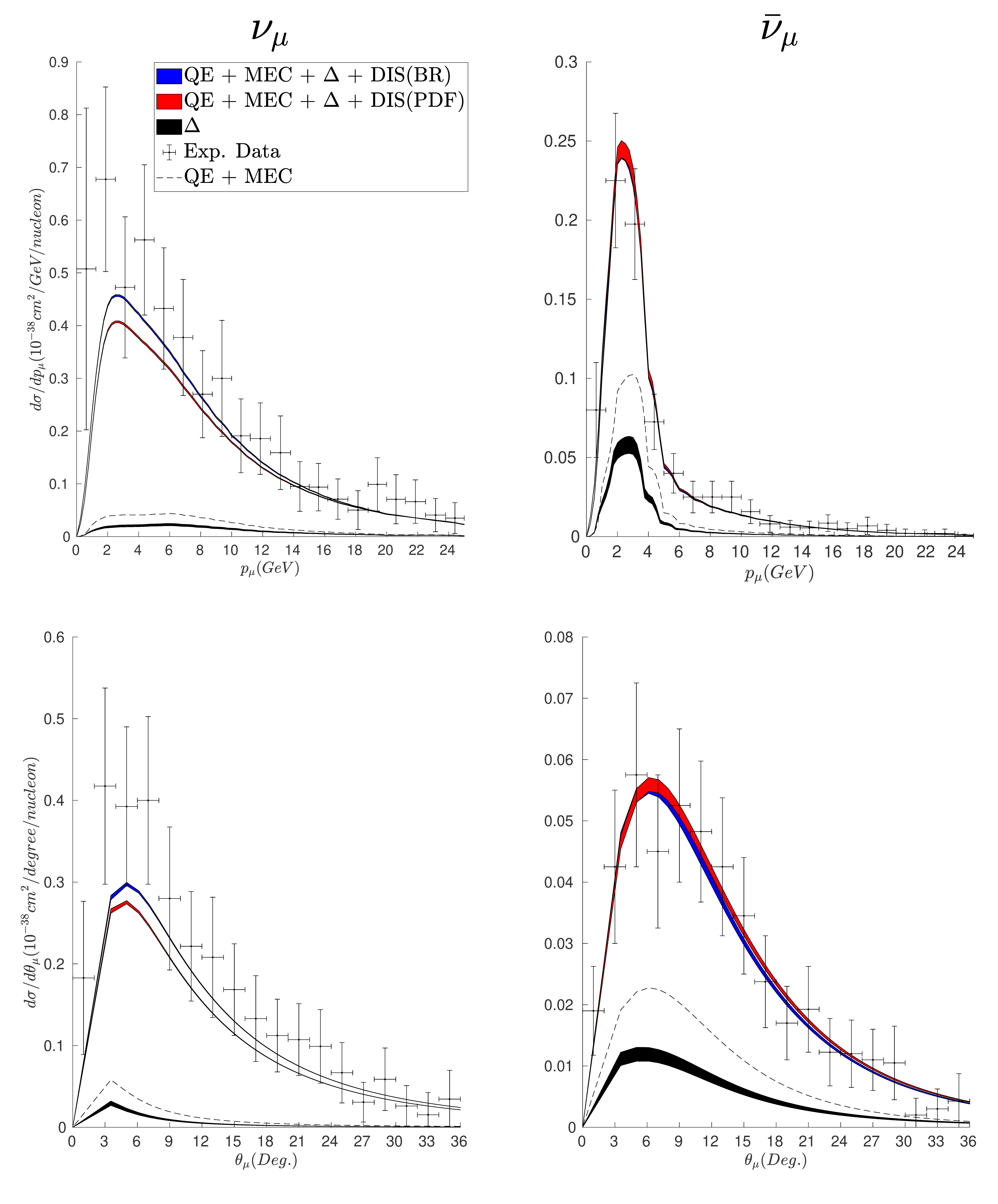}
    \caption{The CC-inclusive ArgoNEUT flux-integrated $\nu_{\mu}$($\bar{\nu}_{\mu}$)-$^{40}$Ar single differential cross section per argon nucleon, displayed as function of the muon momentum (top panels) or the muon scattering angle (bottom).  Legend as in previous figures (see Table \ref{TableContribution}). Data taken from \cite{Acciarri_2014}. 
\label{ArgoNEUT_new} }
\end{figure*}

As for MicroBooNE, in the ArgoNEUT experiment the target is liquid argon, but the neutrino (antineutrino) beam flux is peaked at much larger values. Results are presented in 
figs.~\ref{ArgoNEUT_old} and \ref{ArgoNEUT_new} where two different neutrino energy fluxes have been considered.


In fig.~\ref{ArgoNEUT_old} we show the CC-inclusive $\nu_{\mu}-^{40}$Ar single-differential cross section per nucleon displayed in function of the muon momentum (left), $p_{\mu}$, and the scattering angle (right), $\theta_{\mu}$, folded with the ArgoNEUT flux that is peaked at 4.3 GeV. As in all the previous cases, the separate contribution of the different channels is displayed. The total response is presented as the red and blue bands corresponding to two different descriptions of the inelastic nucleon structure functions: BR (blue band) and PDF (red band). As observed, the BR prediction is larger. This is in contrast with the results presented in the previous cases, and it is probably connected with the very different neutrino energy flux involved in ArgoNEUT. We do not show results for the BC parameterization as it has been fitted within a kinematical range, $0\leq Q^{2} \leq 8$ GeV$^{2}$, much smaller than the $Q^{2}$-values involved in this analysis.

In fig.~\ref{ArgoNEUT_new} we show the cross sections but with the neutrino (antineutrino) beam flux peaked at 9.6 (3.6) GeV. In the case of neutrinos (left panels) the BR parameterization (blue band) leads to significantly larger cross sections. This is in accordance with the results in fig.~\ref{ArgoNEUT_old}, although here the relative discrepancy between the two parameterizations has increased. This is connected with the much larger value of the neutrino energy where the flux is peaked, namely 9.6 GeV versus 4.3 GeV.
Note also the significant difference in the shape of the neutrino cross sections for the two neutrino fluxes. On the contrary, there is almost no difference between the predictions of the two parameterizations in the case of antineutrinos (right panels).


Regarding comparison with the data, we observe that the models underestimate neutrino cross sections, while they agree nicely with antineutrinos. A basic difference between both cases comes from the relative contribution of the QE, MEC and $\Delta$. Whereas these channels contribute significantly to the cross section (similarly to the DIS) in the case of antineutrinos, being their role essential to explain the data, the opposite occurs for neutrinos, particularly with the neutrino beam flux peaked at larger values (fig.~\ref{ArgoNEUT_new}). Here, the relative contribution of QE, MEC and $\Delta$ is very small, of the order of $15\%$ or less, compared with the DIS response. In the case of the neutrino flux peaked at smaller values (fig.~\ref{ArgoNEUT_old}), the relative contribution of QE, MEC and $\Delta$ channels increases slightly. These results are consistent with the ones obtained for MINERvA where the QE, MEC and $\Delta$ channels were shown to play a very minor role compared with the DIS at some kinematics. However, further analysis is needed to understand these significant discrepancies between theory and data for neutrinos and antineutrinos. Different models for $\Delta$ contributions and further analysis for inelastic contribution are presently in progress. 

\section{Conclusions \label{Conclusions}}

A theoretical model capable of reproducing electron scattering data across the whole energy spectrum, from the quasielastic (QE) region up to deep inelastic scattering (DIS), was developed by our group in Ref.~\cite{Megias_2016}. This model is based on scaling/superscaling and incorporates the role of two-particle two-hole excitations (2p-2h), $\Delta$ and heavier nucleon resonances, as well as the region of very high inelasticity (DIS). Theoretical predictions have been shown to reproduce with high precision electron scattering data for a large variety of kinematical situations. 

The superscaling model was also extended to the weak interaction, but only taking into account the QE,  2p-2h and $\Delta$ resonance channels.
In this work we have included in the theoretical description new ingredients associated with heavier nucleon resonances and deep inelastic scattering. This is not straightforward as the axial character of the weak interaction introduces a new inelastic single-nucleon structure function, $W_3$, which must be modeled relying on the quark/parton model or other assumptions. Different options have been explored in this work. On the one hand, we have used different parameterizations considered in the literature, {\it i.e.,} Bodek-Ritchie (BR) and Bosted-Christy (BC), not only for the inelastic electromagnetic structure functions $W_1$ and $W_2$, but also for $W_3$, using a relation that connects $W_3$ to $W_2$ and the antiquark distribution. On the other hand, the responses have been calculated using Parton Distribution Functions (PDF). Whereas the former SuSAv2-$\Delta$ approach is more appropriate at intermediate-high energies , that is, the region where the nucleon resonances are located (up to invariant masses of about 2 GeV), the PDF description works much better at the largest energy, the DIS region, while it clearly fails at lower values.

The models have been first applied to electron scattering providing in general a very good description of data at very different kinematics. Then, a systematic analysis of weak processes has been performed by comparing our predictions with available data for charged current muon neutrino-nucleus reactions. The models have been applied to the study of different experiments: T2K, MINERvA, MicroBooNE, and ArgoNEUT. These involve carbon and argon as nuclear targets. It has been shown that, as the neutrino energy fluxes differ significantly, the relative contribution of the different channels, QE, 2p-2h, $\Delta$ and DIS strongly depends on the experiment. 

For T2K, the channel that dominates is the QE one, and the data in most of the kinematical situations are well described by the models. Furthermore, at forward angles the contribution of DIS gets larger, being crucial to explain the experiment. Similar comments also apply to MicroBooNE, although here some discrepancies between data and theoretical predictions are observed at both backward and forward angles. This is consistent with previous studies based on Monte Carlo analyses and other theoretical calculations.

The situation is different for the higher energy experiments MINERvA and ArgoNEUT, where the DIS contribution dominates in most of the kinematics explored. Here, the theoretical predictions are clearly below the data in the region where the cross sections reach their maxima. This is strictly true in the case of neutrinos for both MINERvA and ArgoNEUT. However, the models provide an excellent description of ArgoNEUT data for antineutrinos. This can be connected with the antineutrino energy flux, rather different from the neutrino ones (peaked at significantly larger values)  and the particular role played by the different channels. Whereas all of them give sizeable (and not so different) contributions for antineutrinos, the DIS channel largely dominates for neutrinos.

This work should be considered as a first step in the description of neutrino-nucleus scattering including the energy spectrum ranging from QE up to DIS. This is crucial to analyze neutrino oscillation experiments where the broad neutrino energy fluxes require knowledge of the contribution of the different reaction channels. In this work, several approaches to the problem, particularly concerning the axial $W_3$ inelastic function, have been explored. The present study shows clearly the applicability of these approaches to describe weak processes, but also their limitations. Although further studies are needed with new models implemented, like the dynamical coupled channels model (DCC)~\cite{PhysRevD.92.074024},  we believe that this work can provide helpful information 
for the analyses of  present and future experiments on neutrino oscillations.




\begin{acknowledgments}
This work is part of the I+D+i project with Ref. PID2020-114687GB-I00, funded by MCIN/AEI/10.13039/501100011033, by the Spanish Ministerio de Ciencia, Innovaci\'on y Universidades and ERDF (European Regional Development Fund) under contracts FIS2017-88410-P and by the Junta de Andalucia (grants No. FQM160, SOMM17/6105/UGR and P$20\_01247$) (J.A.C., GDM, JGR); it is supported in part by the University of Tokyo ICRR's Inter-University Research Program FY2020 $\&$ FY2021 (J.A.C., M.B.B., J.G.R., G.D.M.), by the European Union's Horizon 2020 research and innovation programme under the Marie Sklodowska-Curie grant agreement No. 839481 (G.D.M.), by the Project BARM-RILO-20 of University of Turin and from INFN, National Project NUCSYS (M.B.B.). 
\end{acknowledgments}

\bibliography{DIS.bib}

\end{document}